\documentclass[10pt]{article}
\usepackage{multicol}
\usepackage{graphicx}
\usepackage{amsmath}
\usepackage[a4paper]{geometry}
\usepackage{hyperref}
\usepackage{rotating}

\begin{document}

\begin{center}
{\Large\bf Kinetic freeze-out properties from transverse momentum
spectra of pions in high energy proton-proton collisions}

\vskip.75cm

Li-Li~Li{\footnote{E-mail: shanxi\_lll@163.com;
shanxi-lll@qq.com}}, Fu-Hu~Liu{\footnote{Correspondence E-mail:
fuhuliu@163.com; fuhuliu@sxu.edu.cn}}

\vskip.25cm

{\small\it Institute of Theoretical Physics \& State Key
Laboratory of Quantum Optics and Quantum Optics Devices,\\ Shanxi
University, Taiyuan, Shanxi 030006, China}

\end{center}

\vskip.5cm

{\bf Abstract:} Transverse momentum spectra of negative and
positive pions produced at mid-(pseudo)rapidity in inelastic or
non-single-diffractive proton-proton collisions over a
center-of-mass energy, $\sqrt{s}$, range from a few GeV to above
10 TeV are analyzed by the blast-wave fit with Boltzmann (Tsallis)
distribution. The blast-wave fit results are well fitting to the
experimental data measured by several collaborations. In a
particular superposition with Hagedorn function, both the
excitation functions of kinetic freeze-out temperature ($T_0$) of
emission source and transverse flow velocity ($\beta_T$) of
produced particles obtained from a given selection in the
blast-wave fit with Boltzmann distribution have a hill at
$\sqrt{s}\approx10$ GeV, a drop at dozens of GeV, and then an
increase from dozens of GeV to above 10 TeV. However, both the
excitation functions of $T_0$ and $\beta_T$ obtained in the
blast-wave fit with Tsallis distribution do not show such a
complex structure, but a very low hill. In another selection for
the parameters or in the superposition with the usual step
function, $T_0$ and $\beta_T$ increase generally quickly from a
few GeV to about 10 GeV and then slightly at above 10 GeV, there
is no such the complex structure, when also studying
nucleus-nucleus collisions.
\\

\\


\vskip1.0cm

\begin{multicols}{2}

{\section{Introduction}}

Chemical and thermal or kinetic freeze-outs are two of important
stages of system evolution in high energy collisions. The
excitation degrees of interacting system at the two stages are
possibly different from each other. To describe different
excitation degrees of interacting system at the two stages, one
can use chemical and kinetic freeze-out temperatures respectively.
Generally, at the stage of chemical freeze-out, the ratios of
different types of particles are no longer changed, and the
chemical freeze-out temperature can be obtained from the ratios of
different particles in the framework of thermal
model~\cite{1}--\cite{3}. At the stage of kinetic freeze-out, the
transverse momentum spectra of different particles are no longer
changed, and the dissociation temperature~\cite{4} or kinetic
freeze-out temperature can be obtained from the transverse
momentum spectra according to the hydrodynamical model~\cite{4}
and the subsequent blast-wave fit with Boltzmann
distribution~\cite{5}--\cite{7} or with Tsallis
distribution~\cite{8}--\cite{10}.

It should be pointed out that the transverse momentum spectra even
though in narrow range contain both the contributions of random
thermal motion and transverse flow of particles. The random
thermal motion and transverse flow reflect the excitation and
expansion degrees of the interacting system (or emission source)
respectively. To extract the kinetic freeze-out temperature from
transverse momentum spectra, we have to exclude the contribution
of transverse flow, that is, we have to disengage the random
thermal motion and transverse flow. There are more than one
methods to disengage the two issues~\cite{4}. The simplest and
easiest method is to use the blast-wave fit with Boltzmann
distribution~\cite{5}--\cite{7} and with Tsallis
distribution~\cite{8}--\cite{10} to analyze the transverse
momentum spectra, though other method such as the alternative
method~\cite{6}\cite{11}--\cite{17} can obtain similar
results~\cite{18}.

The early blast-wave fit is based on the Boltzmann
distribution~\cite{5}--\cite{7}. An alternative blast-wave fit is
used due to the Tsallis distribution~\cite{8}--\cite{10}. Both
types of blast-wave fit can be used to disengage the random
thermal motion and transverse flow. Then, the kinetic freeze-out
temperature of interacting system and transverse flow velocity of
light flavor particles can be extracted. Most of light flavor
particles are produced in soft excitation process and have narrow
transverse momentum range up to 2$\sim$3 GeV/$c$. A few part of
light flavor particles are produced in hard scattering process and
have higher transverse momenta. Generally, heavy flavor particles
are produced via hard scattering process. From the point of view
of disengaging or extraction, particles produced in hard
scattering process are not needed to consider by us.

The excitation function of kinetic freeze-out temperature, that
is, the dependence of kinetic freeze-out temperature on collision
energy, is very interesting for us to study the properties of high
energy collisions. We think that the particular change of
excitation functions of kinetic freeze-out temperature and other
parameters are related to the critical-end-point (CEP) of phase
transition from hadronic matter to quark-gluon plasma (QGP or
quark matter) happened in central nucleus-nucleus ($AA$ or
$A$-$A$) collisions, where the particular change means the
appearances of saturation, minimum, maximum, knee point,
asymptotical line, etc. For high multiplicity proton-proton ($pp$
or $p$-$p$) collisions, even for minimum-bias $pp$ collisions, the
particular change of excitation functions of some parameters are
expected to compare with those in $AA$ collisions, where the quark
degree of freedom in minimum-bias $pp$ collisions is expected to
play initially a main role at the energy of particular change,
though QGP is not expected to form in minimum-bias $pp$ collisions
due to small system and products. The minimum of excitation
function is also related to the soft point of equation of state
(EOS) of hadronic matter or QGP, which is also related to the
phase transition.

Although there are many studies on the excitation functions of
kinetic freeze-out temperature and other parameters, the results
seem to be inconsistent. For example, over a center-of-mass energy
per nucleon pair, $\sqrt{s_{NN}}$, range from a few GeV to a few
TeV, the excitation function of kinetic freeze-out temperature in
gold-gold (Au-Au) and lead-lead (Pb-Pb) collisions initially
increases and then inconsistently saturates~\cite{19}\cite{20},
increases~\cite{21}, or decreases~\cite{22}\cite{23} with the
increase of collision energy. On the contrary, the excitation
function of the chemical freeze-out temperature shows initially
increases and then consistently saturates with collision
energy~\cite{1}--\cite{4}. Comparatively, as the basic processes
in $AA$ collisions, $pp$ collisions are very important in the
study of the mentioned excitation functions. However, the
excitation functions in $pp$ collisions are short of studies. We
hope to study the particular changes of excitation functions in
$pp$ collisions due to they being also related to quark degree of
freedom, but not QGP. Indeed, it is worth to study the excitation
functions of kinetic freeze-out temperature and other parameters
in $pp$ collisions and to judge their tendencies over an energy
range from GeV to TeV.

In this paper, by using the blast-wave fit with Boltzmann
distribution~\cite{5}--\cite{7} and with Tsallis
distribution~\cite{8}--\cite{10}, we study the excitation
functions of some concerned quantities in inelastic (INEL) or
non-single-diffractive (NSD) $pp$ collisions which are closer to
peripheral nuclear collisions comparing with central ones. The
experimental transverse momentum spectra of negative and positive
pions ($\pi^-$ and $\pi^+$) measured mainly at the mid-rapidity by
the NA61/SHINE Collaboration~\cite{24} at the the Super Proton
Synchrotron (SPS) and its beam energy scan (BES) program, the
PHENIX~\cite{25} and STAR~\cite{6} Collaborations at the
Relativistic Heavy Ion Collider (RHIC), as well as the
ALICE~\cite{26} and CMS~\cite{27}\cite{28} Collaborations at the
Large Hadron Collider (LHC) are analyzed, while the data in the
forward and backward rapidity regions are not available in most
cases.

The remainder of this paper is structured as follows. The
formalism and method are shortly described in Section 2. Results
and discussion are given in Section 3. In Section 4, we summarize
our main observations and conclusions.
\\

{\section{Formalism and method}}

There are two main processes of particle productions, namely the
soft excitation process and the hard scattering process, in high
energy collisions. For the soft excitation process, the method
used in the present work is the blast-wave fit~\cite{5}--\cite{10}
that has wide applications in particle productions. The blast-wave
fit is based on two types of distributions. One is the Boltzmann
distribution~\cite{5}--\cite{7} and another one is the Tsallis
distribution~\cite{8}--\cite{10}. As an application of the
blast-wave fit, we present directly its formalisms in the
following. Although the blast-wave fit has abundant connotations,
we focus only our attention on the formalism of transverse
momentum ($p_T$) distribution in which the kinetic freeze-out
temperature ($T_0$) and mean transverse flow velocity ($\beta_T$)
are included.

We are interested in the blast-wave fit with Boltzmann
distribution in its original form. According to
refs.~\cite{5}--\cite{7}, the blast-wave fit with Boltzmann
distribution results in the probability density distribution of
$p_T$ to be
\begin{align}
f_1(p_T)= & \frac{1}{N}\frac{dN}{dp_T}= C_1 p_T m_T \int_0^R rdr \times \nonumber\\
& I_0 \bigg[\frac{p_T \sinh(\rho)}{T_0} \bigg] K_1 \bigg[\frac{m_T
\cosh(\rho)}{T_0} \bigg],
\end{align}
where $C_1$ is the normalized constant, $m_T=\sqrt{p_T^2+m_0^2}$
is the transverse mass, $m_0$ is the rest mass, $r$ is the radial
coordinate in the thermal source, $R$ is the maximum $r$ which can
be regarded as the transverse size of source in the case of
neglecting the expansion, $r/R$ is the relative radial position
which has in fact more meanings than $r$ and $R$ themselves, $I_0$
and $K_1$ are the modified Bessel functions of the first and
second kinds respectively, $\rho= \tanh^{-1} [\beta(r)]$ is the
boost angle, $\beta(r)= \beta_S(r/R)^{n_0}$ is a self-similar flow
profile, $\beta_S$ is the flow velocity on the surface, and
$n_0=2$ is used in the original form~\cite{5}. There is the
relation between $\beta_T$ and $\beta(r)$. As a mean of
$\beta(r)$, $\beta_T=(2/R^2)\int_0^R r\beta(r)dr =
2\beta_S/(n_0+2)$.

We are also interested in the blast-wave fit with Tsallis
distribution in its original form. According to
ref.~\cite{8}--\cite{10}, the blast-wave fit with Tsallis
distribution results in the $p_T$ distribution to be
\begin{align}
f_2(p_T)= & \frac{1}{N}\frac{dN}{dp_T}= C_2 p_T m_T \int_{-\pi}^{\pi} d\phi \int_0^R rdr \Big\{1+ \nonumber\\
& \frac{q-1}{T_0} \big[ m_T \cosh(\rho) -p_T \sinh(\rho)
\cos(\phi)\big] \Big\}^{-1/(q-1)},
\end{align}
where $C_2$ is the normalized constant, $q$ is an entropy index
that characterizes the degree of non-equilibrium, $\phi$ denotes
the azimuthal angle, and $n_0=1$ is used in the original
form~\cite{8}. Because of $n_0$ being an insensitive quantity, the
results corresponding to $n_0=1$ and 2 for the blast-wave fit with
Boltzmann or Tsallis distribution are harmonious~\cite{18}. In
fact, in some literature~\cite{28a}, $n_0$ is regarded as a free
parameter which changes largely by several times and increases 1
in the number of free parameter, which is not our expectation in
the present work. In addition, the index $-1/(q-1)$ used in Eq.
(2) can be replaced by $-q/(q-1)$ due to $q$ being very close to
1. This substitution results in a small and negligible difference
in the Tsallis distribution~\cite{29}\cite{30}.

For a not too wide $p_T$ spectrum, the above two equations can be
used to describe the $p_T$ spectrum and to extract $T_0$ and
$\beta_T$. For a wide $p_T$ spectrum, we have to consider the
contribution of hard scattering process. According to the quantum
chromodynamics (QCD) calculus~\cite{31}--\cite{33}, the
contribution of hard scattering process is parameterized to be an
inverse power-law
\begin{align}
f_H(p_T)=\frac{1}{N}\frac{dN}{dp_T}=Ap_T \bigg( 1+\frac{p_T}{ p_0}
\bigg)^{-n}
\end{align}
which is the Hagedorn function~\cite{34}\cite{35}, where $p_0$ and
$n$ are free parameters, and $A$ is the normalization constant
related to the free parameters. In
literature~\cite{36},~\cite{37}--\cite{41}, and~\cite{42}, there
are respectively modified Hagedorn functions
\begin{align}
f_H(p_T)=\frac{1}{N}\frac{dN}{dp_T}=A \frac{p^2_T}{m_T} \bigg(
1+\frac{p_T}{ p_0} \bigg)^{-n},
\end{align}
\begin{align}
f_H(p_T)=\frac{1}{N}\frac{dN}{dp_T}=Ap_T \bigg( 1+\frac{p^2_T}{
p^2_0} \bigg)^{-n},
\end{align}
and
\begin{align}
f_H(p_T)=\frac{1}{N}\frac{dN}{dp_T}=A \bigg( 1+\frac{p^2_T}{
p^2_0} \bigg)^{-n},
\end{align}
where the three normalization constants $A$, free parameters
$p_0$, and free parameters $n$ are severally different, though the
same symbols are used to avoid trivial expression.

The experimental $p_T$ spectrum distributed in a wide range can be
described by a superposition of the contributions of the soft
excitation and hard scattering processes. If one of Eqs. (1) and
(2) describes the contribution of the soft excitation process, one
of Eqs. (3)--(6) describes the contribution of the hard scattering
process. To describe the spectrum in a wide $p_T$ range, we can
superpose a two-component superposition like this
\begin{align}
f_0(p_T)=kf_S(p_T)+(1-k)f_H(p_T),
\end{align}
where $k$ ($1-k$) denotes the contribution fraction of the soft
excitation (hard scattering) process, and $f_S(p_T)$ denotes one
of Eqs. (1) and (2). As for the four $f_H(p_T)$, we are inclined
to the first one due to its more applications. Naturally, we have
the normalization condition $\int_0^{p_{T\max}}f_0(p_T)dp_T=1$,
where $p_{T\max}$ denotes the maximum $p_T$. In Eq. (7), the soft
component contributes in low $p_T$ region and the hard component
contributes in whole $p_T$ range. The two contributions overlap
each other in low $p_T$ region.

According to Hagedorn's model~\cite{34}, we may also use the usual
step function to superpose the two functions. That is
\begin{align}
f_0(p_T)=A_1\theta(p_1-p_T)f_S(p_T)+A_2\theta(p_T-p_1)f_H(p_T),
\end{align}
where $A_1$ and $A_2$ are constants which result in the two
components to be equal to each other at $p_T=p_1 \approx 2\sim3$
GeV/$c$. The contribution fraction of the soft excitation (hard
excitation) process in Eq. (8) is $k=\int_0^{p_1} A_1f_S(p_T)dp_T$
[$1-k=\int_{p_1}^{p_{T\max}} A_2f_H(p_T)dp_T$] due to
$\int_0^{p_{T\max}}f_0(p_T)dp_T=1$. In Eq. (8), the soft component
contributes in low $p_T$ region and the hard component contributes
in high $p_T$ region. The two contributions link with each other
at $p_T=p_1$.

In some cases, the contribution of resonance production for pions
in very-low $p_T$ range has to be considered. We can use a
very-soft component for the $p_T$ range from 0 to 0.2$\sim$0.3
GeV/$c$ which covers the contribution of resonance production. Let
$k_{VS}$ and $k_S$ denote the contribution fractions of the
very-soft and soft processes respectively. Eq. (7) is revised to
\begin{align}
f_0(p_T)=& k_{VS}f_{VS}(p_T) +k_Sf_S(p_T) \nonumber\\
& +(1-k_{VS}-k_S)f_H(p_T),
\end{align}
where $f_{VS}(p_T)$ denotes one of Eqs. (1) and (2) as $f_S(p_T)$,
but having smaller parameter values comparing with $f_S(p_T)$.
Anyhow, both the very-soft and soft processes are belong to the
soft process. Correspondingly, Eq. (8) is revised to
\begin{align}
f_0(p_T)=&A_1 \theta(p_1-p_T) f_{VS}(p_T)  \nonumber\\
&+ A_2\theta(p_T-p_1)\theta(p_2-p_T)f_S(p_T) \nonumber\\
&+A_3\theta(p_T-p_2)f_H(p_T),
\end{align}
where $A_1$, $A_2$, and $A_3$ are constants which result in the
two contiguous components to be equal to each other at $p_T=p_1$
and $p_T=p_2$.

The above two types of superpositions [Eqs. (7) and (8)] treat the
soft and hard components by different ways in the whole $p_T$
range. Eq. (7) means that the soft component contributes in a
range from 0 up to 2$\sim$3 GeV/$c$ or a little more. The hard
component contributes in the whole $p_T$ range, though the main
contributor in the low $p_T$ region is the soft component and the
sole contributor in the high $p_T$ region is the hard component.
Eq. (8) shows that the soft component contributes in a range from
0 up to $p_1$, and the hard component contributes in a range from
$p_1$ up to the maximum. The boundary of the contributions of soft
and hard components is $p_1$. There is no mixed region for the two
components in Eq. (8).

In the case of including only the soft component, Eqs. (7) and (8)
are the same. In the case of including both the soft and hard
components, their common parameters such as $T_0$, $\beta_T$,
$p_0$, and $n$ should be severally to have small differences from
each other. To avoid large differences, we should select the
experimental data in a narrow $p_T$ range. In addition, the
very-soft component in Eqs. (9) and (10) does not need to consider
in some cases due to the fact that the spectrum in very-low $p_T$
range are possibly not available in experiment. Then, Eqs. (9) and
(10) are degenerated to Eqs. (7) and (8) respectively in some
cases.

In the fit process, firstly, we use Eq. (7) to extract the related
parameters, where $f_S(p_T)$ and $f_H(p_T)$ are exactly Eqs. (1)
or (2) and (3) respectively. Regardless of Eq. (1) or (2) is
regarded as $f_S(p_T)$, the situation is similar due to Eqs. (1)
and (2) being harmonious in trends of parameters~\cite{18}, though
one more parameter (the entropy index $q$) is needed in Eq. (2).
Secondly, we use Eq. (8) to extract the related parameters as
comparisons with those from Eqs. (7). In the case of Eqs. (7) and
(8) being not suitable, Eqs. (9) and (10) can be used due to the
very-low component from resonance decays being included in the
first items of the two fits. Meanwhile, the contribution of
resonance decays in the low component is naturally included in the
second items of Eqs. (9) and (10) or in the first items of Eqs.
(7) and (8). Thus, the contribution of resonance decays which is
available in the very-low or low component in experiment is
naturally considered by us.

It should be pointed out that although Eqs. (9) and (10) are not
used in the final fits in the present work, we hope to keep them
here due to the fact that we need them to give further
explanations. From theses explanations, one can see the relations
between Eqs. (9)/(10) and (7)/(8), as well as possible
applications of Eqs. (9) and (10).
\\

\begin{figure*}[!htb]
\begin{center}
\includegraphics[width=16.0cm]{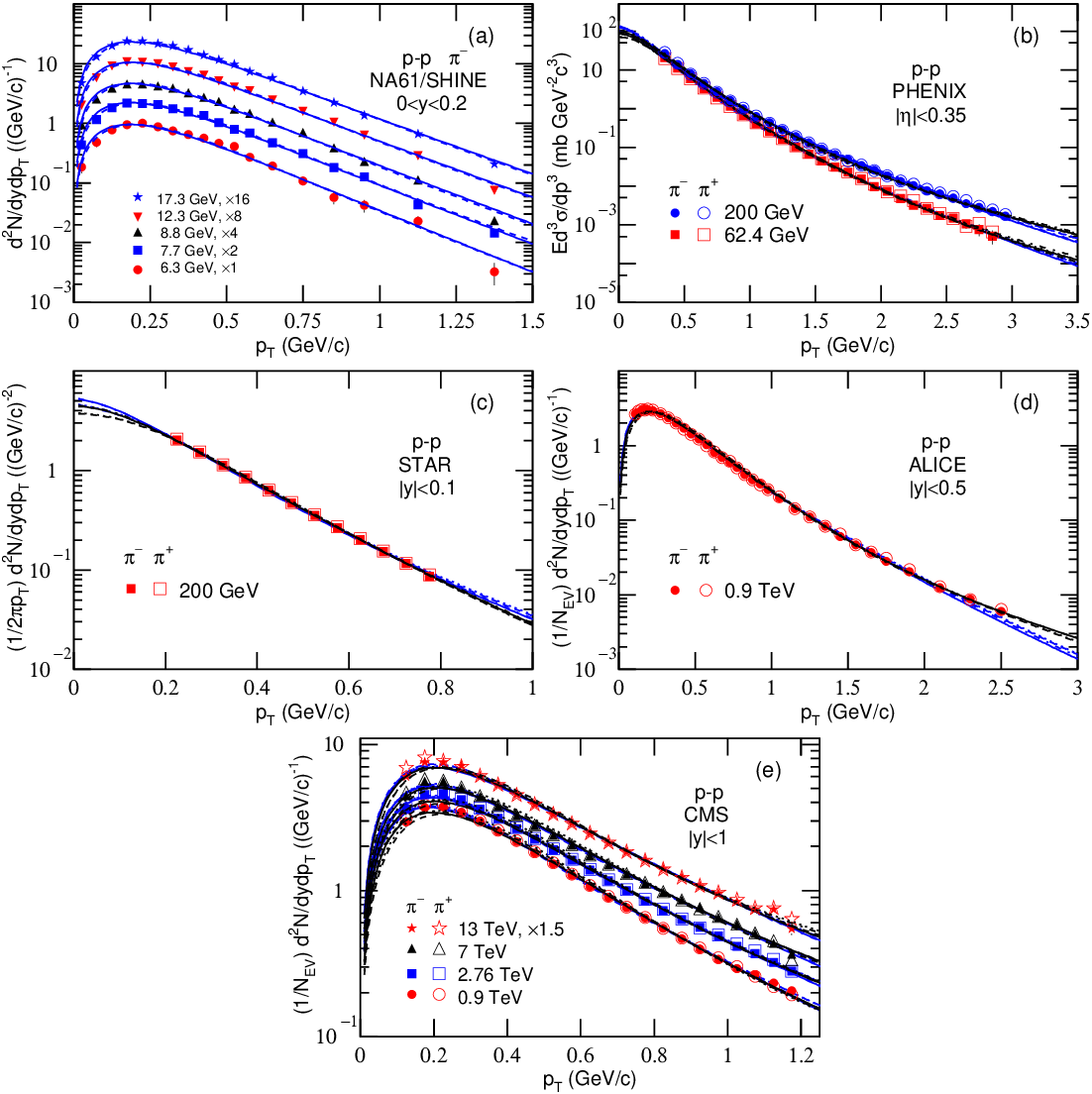}
\end{center}
{\small Fig. 1. Transverse momentum spectra of $\pi^-$ and $\pi^+$
produced at mid-(pseudo)rapidity in $pp$ collisions at high
energies, where the mid-(pseudo)rapidity intervals and energies
are marked in the panels. The symbols presented in panels (a)--(e)
represent the data of NA61/SHINE~\cite{24}, PHENIX~\cite{25},
STAR~\cite{6}, ALICE~\cite{26}, and CMS~\cite{27}\cite{28}
Collaborations, respectively, where in panel (a) only the spectra
of $\pi^-$ are available, and panel (c) is for NSD events and
other panels are for INEL events. Although the $pp$ collisions are
divided on the multiplicity classes in experiments on the LHC, we
have used the minimum-bias INEL events~\cite{26}--\cite{28}. In
some cases, different amounts marked in the panels are used to
scale the data for clarity. The blue solid (dotted) curves are our
results for $\pi^-$ ($\pi^+$) spectra fitted by Eq. (7) through
Eqs. (1) and (3), and the blue dashed (dot-dashed) curves are our
results for $\pi^-$ ($\pi^+$) spectra fitted by Eq. (7) through
Eqs. (2) and (3), by the first set of parameter values. The
results by the second set of parameters (if available) are
presented by the black curves.}
\end{figure*}

{\section{Results and discussion}}

{\subsection {Comparison with data by Eq. (7)}}

Figure 1 shows the transverse momentum spectra of $\pi^-$ and
$\pi^+$ produced at mid-(pseudo)rapidity in $pp$ collisions at
high center-of-mass energies, where different mid-(pseudo)rapidity
($y$ or $\eta$) intervals and energies ($\sqrt{s}$) are marked in
the panels. Different forms of the spectra are used due to
different Collaborations, where $N$, $E$, $p$, $\sigma$, and
$N_{EV}$ denote the particle number, energy, momentum,
cross-section, and event number, respectively. The closed and open
symbols presented in panels (a)--(e) represent the data of $\pi^-$
and $\pi^+$ measured by the NA61/SHINE~\cite{24},
PHENIX~\cite{25}, STAR~\cite{6}, ALICE~\cite{26}, and
CMS~\cite{27}\cite{28} Collaborations, respectively, where in
panel (a) only the spectra of $\pi^-$ are available, and panel (c)
is for NSD events and other panels are for INEL events. Although
the $pp$ collisions are divided on the multiplicity classes in
experiments on the LHC, we have used the minimum-bias INEL
events~\cite{26}--\cite{28} which can be regarded as the
combination of the events in different multiplicity classes with
different yields. We regret not using the ATLAS Collaboration
results~\cite{42a}--\cite{42e} given no data on the $p_T$ spectra
of $\pi^-$ and $\pi^+$. These data show wider $p_T$ spectra of
charged particles to be used in our foreseen studies. In some
cases, different amounts marked in the panels are used to scale
the data for clarity. We have fitted the data by two sets of
parameter values in Eq. (7) so that we can see the fluctuations of
parameter values. The blue solid (dotted) curves are our results
for $\pi^-$ ($\pi^+$) spectra fitted by Eq. (7) through Eqs. (1)
and (3), and the blue dashed (dot-dashed) curves are our results
for $\pi^-$ ($\pi^+$) spectra fitted by Eq. (7) through Eqs. (2)
and (3), by the first set of parameter values, in which $k$ is
taken to close to 0.5 as much as possible. The black curves are
our results fitted by the second set of parameter values for
comparison, in which $k$ is taken to close to 1 as much as
possible. The values of free parameters ($T_0$, $\beta_T$, $q$ if
available, $k$, $p_0$, and $n$), normalization constant ($N_0$),
$\chi^2$, and number of degrees of freedom (ndof) corresponding to
the curves in Fig. 1 are listed in Tables 1 and 2, where the
errors of fit parameters are obtained by the statistical
simulation method~\cite{42f}, no matter what $\chi^2$/ndof is. The
parameter values presented in terms of value$_1$/value$_2$ denote
respectively the first and second sets of parameter values in Eq.
(7) through Eqs. (1) [Eqs. (2)] and (3) in which $k\neq1$. One can
see that Eq. (7) with two sets of parameter values describes the
$p_T$ spectra at mid-(pseudo)rapidity in $pp$ collisions over an
energy range from a few GeV to above 10 TeV. The blast-wave fit
with Boltzmann distribution and with Tsallis distribution presents
similar results. The free parameters show some laws in the
considered energy range. For a given parameter, its fluctuation at
given energy is obvious in some cases. Because of the data being
not available in very-low $p_T$ range, Eq. (9) is not used in the
fit.

\clearpage

\end{multicols}
\begin{sidewaystable}
{\scriptsize Table 1. Values of free parameters ($T_0$, $\beta_T$,
$k$, $p_0$, and $n$), normalization constant ($N_0$), $\chi^2$,
and ndof corresponding to the solid (dotted) curves for $\pi^-$
($\pi^+$) spectra in Fig. 1 in which different data are measured
in different mid-(pseudo)rapidity intervals at different energies
by different Collaborations. The values presented in terms of
value$_1$/value$_2$ denote respectively the first and second sets
of parameter values in Eq. (7) through Eqs. (1) and (3) in which
$k\neq 1$.} \vspace{-4mm}
\begin{center}
{\tiny
\begin{tabular} {ccccccccccc}\\ \hline\hline Collab. & $\sqrt{s}$ (GeV) &
Part. & $T_0$ (MeV) & $\beta_T$ ($c$) & $k$ & $p_0$ (GeV/c) & $n$ & $N_0$ & $\chi^2$ & ndof \\
\hline
NA61/  & 6.3   & $\pi^-$ & $108\pm5$ & $0.30\pm0.02$ & 1 & $-$ & $-$ & $0.08\pm0.01$ & 21 & 15\\
SHINE  & 7.7   & $\pi^-$ & $109\pm5$ & $0.31\pm0.02$ & 1 & $-$ & $-$ & $0.10\pm0.01$ & 34 & 15\\
INEL   & 8.8   & $\pi^-$ & $110\pm5$ & $0.31\pm0.02$ & 1 & $-$ & $-$ & $0.10\pm0.01$ & 73 & 15\\
       & 12.3  & $\pi^-$ & $111\pm6$ & $0.32\pm0.02$ & 1 & $-$ & $-$ & $0.12\pm0.01$ & 59 & 15\\
       & 17.3  & $\pi^-$ & $112\pm6$ & $0.33\pm0.02$ & 1 & $-$ & $-$ & $0.13\pm0.01$ & 26 & 15\\
\hline
PHENIX & 62.4  & $\pi^-$ & $96\pm5/114\pm6$  & $0.27\pm0.01/0.34\pm0.02$ & $0.66\pm0.01/0.98\pm0.01$ & $3.60\pm0.18/6.06\pm0.30$ & $19.23\pm0.96/18.63\pm0.93$ & $21.55\pm1.08/18.96\pm0.95$ & 7/28  & 20\\
INEL   &       & $\pi^+$ & $96\pm5/114\pm6$  & $0.27\pm0.01/0.34\pm0.02$ & $0.66\pm0.01/0.98\pm0.01$ & $3.63\pm0.18/6.07\pm0.30$ & $19.03\pm0.95/18.63\pm0.93$ & $20.81\pm1.04/18.57\pm0.93$ & 11/11 & 20\\
       & 200   & $\pi^-$ & $100\pm5/116\pm5$ & $0.30\pm0.02/0.36\pm0.02$ & $0.62\pm0.01/0.96\pm0.02$ & $4.19\pm0.21/6.45\pm0.32$ & $19.01\pm0.95/18.01\pm0.90$ & $23.98\pm1.20/22.77\pm1.14$ & 26/21 & 21\\
       &       & $\pi^+$ & $100\pm5/115\pm5$ & $0.30\pm0.02/0.35\pm0.02$ & $0.62\pm0.01/0.96\pm0.02$ & $4.21\pm0.21/6.46\pm0.32$ & $19.01\pm0.95/18.00\pm0.90$ & $24.41\pm1.22/24.41\pm1.22$ & 54/26 & 21\\
\hline
STAR   & 200   & $\pi^-$ & $98\pm6/112\pm5$  & $0.29\pm0.02/0.34\pm0.02$ & $0.61\pm0.03/0.98\pm0.02$ & $4.01\pm0.20/6.00\pm0.30$ & $19.21\pm0.96/18.61\pm0.93$ & $0.27\pm0.01/0.92\pm0.05$ & 76/3 & 6\\
NSD    &       & $\pi^+$ & $99\pm6/112\pm5$  & $0.29\pm0.02/0.34\pm0.02$ & $0.60\pm0.03/0.98\pm0.02$ & $4.01\pm0.20/6.00\pm0.30$ & $19.21\pm0.96/18.61\pm0.93$ & $0.27\pm0.01/0.27\pm0.01$ & 96/4 & 6\\
\hline
ALICE  & 900   & $\pi^-$ & $101\pm5/116\pm6$ & $0.31\pm0.02/0.36\pm0.02$ & $0.63\pm0.02/0.94\pm0.02$ & $4.39\pm0.22/6.81\pm0.34$ & $18.89\pm0.94/17.35\pm0.87$ & $1.47\pm0.07/1.47\pm0.07$ & 38/126 & 27\\
INEL   &       & $\pi^+$ & $101\pm5/116\pm6$ & $0.31\pm0.02/0.35\pm0.02$ & $0.63\pm0.02/0.95\pm0.02$ & $4.42\pm0.22/6.96\pm0.35$ & $18.81\pm0.94/17.35\pm0.87$ & $1.47\pm0.07/1.47\pm0.07$ & 49/137 & 27\\
\hline
CMS    & 900   & $\pi^-$ & $101\pm5/115\pm6$ & $0.31\pm0.02/0.35\pm0.02$ & $0.63\pm0.03/0.91\pm0.02$ & $4.43\pm0.22/7.08\pm0.35$ & $18.71\pm0.93/17.13\pm0.87$ & $3.65\pm0.18/3.49\pm0.17$ & 24/62 & 16\\
INEL   &       & $\pi^+$ & $101\pm5/115\pm5$ & $0.31\pm0.02/0.35\pm0.02$ & $0.63\pm0.03/0.92\pm0.02$ & $4.43\pm0.22/7.04\pm0.35$ & $18.71\pm0.93/17.16\pm0.86$ & $3.70\pm0.19/3.55\pm0.18$ & 16/59 & 16\\
       & 2760  & $\pi^-$ & $103\pm6/116\pm4$ & $0.33\pm0.02/0.36\pm0.02$ & $0.63\pm0.03/0.90\pm0.02$ & $4.68\pm0.23/7.80\pm0.39$ & $18.41\pm0.92/16.45\pm0.82$ & $4.47\pm0.22/4.31\pm0.22$ & 34/70 & 16\\
       &       & $\pi^+$ & $102\pm6/116\pm5$ & $0.33\pm0.02/0.36\pm0.02$ & $0.63\pm0.04/0.91\pm0.02$ & $4.69\pm0.23/7.90\pm0.39$ & $18.39\pm0.92/16.35\pm0.82$ & $4.55\pm0.23/4.43\pm0.22$ & 35/74 & 16\\
       & 7000  & $\pi^-$ & $104\pm5/117\pm6$ & $0.34\pm0.02/0.37\pm0.02$ & $0.62\pm0.03/0.89\pm0.02$ & $4.79\pm0.24/8.00\pm0.40$ & $18.21\pm0.91/16.13\pm0.81$ & $5.50\pm0.27/5.49\pm0.27$ & 48/67 & 16\\
       &       & $\pi^+$ & $103\pm5/116\pm4$ & $0.34\pm0.02/0.36\pm0.02$ & $0.61\pm0.04/0.89\pm0.02$ & $4.80\pm0.24/8.20\pm0.41$ & $18.21\pm0.91/16.00\pm0.80$ & $5.54\pm0.28/5.52\pm0.28$ & 55/70 & 16\\
       & 13000 & $\pi^-$ & $105\pm5/117\pm5$ & $0.35\pm0.02/0.37\pm0.02$ & $0.61\pm0.03/0.89\pm0.02$ & $4.90\pm0.24/8.30\pm0.41$ & $18.11\pm0.90/15.99\pm0.80$ & $5.07\pm0.25/5.07\pm0.25$ & 30/28 & 16\\
       &       & $\pi^+$ & $104\pm5/117\pm5$ & $0.34\pm0.02/0.36\pm0.02$ & $0.64\pm0.04/0.88\pm0.02$ & $5.00\pm0.25/8.99\pm0.43$ & $18.00\pm0.90/15.59\pm0.78$ & $5.12\pm0.26/5.15\pm0.26$ & 36/42 & 16\\
\hline
\end{tabular}}
\end{center}

{\scriptsize Table 2. Values of free parameters ($T_0$, $\beta_T$,
$q$ $k$, $p_0$, and $n$), normalization constant ($N_0$),
$\chi^2$, and ndof corresponding to the dashed (dot-dashed) curves
for $\pi^-$ ($\pi^+$) spectra in Fig. 1 in which different data
are measured in different mid-(pseudo)rapidity intervals at
different energies by different Collaborations. The values
presented in terms of value$_1$/value$_2$ denote respectively the
first and second sets of parameter values in Eq. (7) through Eqs.
(2) and (3) in which $k\neq 1$.} \vspace{-4mm}
\begin{center}
{\tiny
\renewcommand\tabcolsep{3.5pt}
\begin{tabular} {cccccccccccc}\\ \hline\hline Collab. & $\sqrt{s}$ (GeV) &
Part. & $T_0$ (MeV) & $\beta_T$ ($c$)& $q$ & $k$ & $p_0$ (GeV/c) & $n$ & $N_0$ & $\chi^2$ & ndof \\
\hline
NA61/  & 6.3   & $\pi^-$ & $81\pm4$ & $0.19\pm0.01$ & $1.05\pm0.002$ & 1 & $-$ & $-$ & $0.08\pm0.01$ & 12 & 14\\
SHINE  & 7.7   & $\pi^-$ & $81\pm4$ & $0.20\pm0.01$ & $1.06\pm0.002$ & 1 & $-$ & $-$ & $0.10\pm0.01$ & 13 & 14\\
INEL   & 8.8   & $\pi^-$ & $83\pm4$ & $0.20\pm0.01$ & $1.05\pm0.002$ & 1 & $-$ & $-$ & $0.10\pm0.01$ & 24 & 14\\
       & 12.3  & $\pi^-$ & $84\pm4$ & $0.21\pm0.01$ & $1.06\pm0.002$ & 1 & $-$ & $-$ & $0.12\pm0.01$ & 14 & 14\\
       & 17.3  & $\pi^-$ & $85\pm4$ & $0.21\pm0.01$ & $1.06\pm0.002$ & 1 & $-$ & $-$ & $0.13\pm0.01$ & 5  & 14\\
\hline
PHENIX & 62.4  & $\pi^-$ & $78\pm4/86\pm4$ & $0.18\pm0.01/0.21\pm0.01$ & $1.04\pm0.01/1.06\pm0.01$ & $0.63\pm0.03/0.98\pm0.05$ & $3.02\pm0.15/5.36\pm0.27$ & $16.99\pm0.85/18.73\pm0.94$ & $19.53\pm0.98/18.26\pm0.91$ & 16/18 & 19\\
INEL   &       & $\pi^+$ & $79\pm5/85\pm4$ & $0.18\pm0.01/0.21\pm0.01$ & $1.04\pm0.01/1.06\pm0.01$ & $0.63\pm0.02/0.97\pm0.05$ & $3.10\pm0.15/5.36\pm0.27$ & $16.99\pm0.85/18.73\pm0.94$ & $18.42\pm0.92/18.26\pm0.91$ & 30/23 & 19\\
       & 200   & $\pi^-$ & $80\pm5/86\pm4$ & $0.19\pm0.01/0.23\pm0.01$ & $1.02\pm0.01/1.06\pm0.01$ & $0.59\pm0.02/0.95\pm0.05$ & $3.53\pm0.17/5.99\pm0.30$ & $16.68\pm0.84/18.23\pm0.91$ & $24.48\pm1.22/23.86\pm1.19$ & 39/24 & 20\\
       &       & $\pi^+$ & $80\pm5/86\pm4$ & $0.19\pm0.01/0.23\pm0.01$ & $1.02\pm0.01/1.06\pm0.01$ & $0.59\pm0.03/0.95\pm0.05$ & $3.53\pm0.17/6.09\pm0.30$ & $16.68\pm0.84/18.23\pm0.91$ & $25.07\pm1.25/24.49\pm1.22$ & 45/68 & 20\\
\hline
STAR   & 200   & $\pi^-$ & $79\pm5/85\pm4$ & $0.19\pm0.01/0.23\pm0.01$ & $1.04\pm0.01/1.05\pm0.01$ & $0.62\pm0.03/0.95\pm0.05$ & $3.70\pm0.18/5.89\pm0.29$ & $16.68\pm0.82/18.43\pm0.92$ & $0.26\pm0.01/0.26\pm0.01$ & 22/39 & 5\\
NSD    &       & $\pi^+$ & $79\pm4/85\pm4$ & $0.19\pm0.01/0.23\pm0.01$ & $1.04\pm0.01/1.05\pm0.01$ & $0.62\pm0.02/0.95\pm0.05$ & $3.70\pm0.18/5.89\pm0.29$ & $16.68\pm0.82/18.43\pm0.92$ & $0.27\pm0.01/0.26\pm0.01$ & 15/28 & 5\\
\hline
ALICE  & 900   & $\pi^-$ & $81\pm5/86\pm4$ & $0.20\pm0.01/0.25\pm0.01$ & $1.03\pm0.01/1.06\pm0.01$ & $0.53\pm0.02/0.93\pm0.05$ & $3.63\pm0.19/6.39\pm0.32$ & $16.68\pm0.81/18.03\pm0.90$ & $1.47\pm0.07/1.47\pm0.07$ & 34/419& 26\\
INEL   &       & $\pi^+$ & $80\pm3/86\pm4$ & $0.20\pm0.01/0.25\pm0.01$ & $1.03\pm0.01/1.06\pm0.01$ & $0.53\pm0.03/0.93\pm0.05$ & $3.64\pm0.19/6.39\pm0.32$ & $16.68\pm0.81/18.03\pm0.90$ & $1.50\pm0.08/1.50\pm0.08$ & 51/558& 26\\
\hline
CMS    & 900   & $\pi^-$ & $81\pm3/87\pm4$ & $0.19\pm0.01/0.25\pm0.01$ & $1.02\pm0.01/1.05\pm0.01$ & $0.51\pm0.03/0.89\pm0.05$ & $3.74\pm0.19/6.79\pm0.34$ & $16.68\pm0.80/17.83\pm0.89$ & $3.67\pm0.18/3.45\pm0.17$ & 8/124 & 15\\
INEL   &       & $\pi^+$ & $80\pm4/87\pm4$ & $0.19\pm0.01/0.25\pm0.01$ & $1.02\pm0.01/1.05\pm0.01$ & $0.51\pm0.02/0.89\pm0.05$ & $3.72\pm0.19/6.79\pm0.34$ & $16.68\pm0.80/17.83\pm0.89$ & $3.74\pm0.19/3.59\pm0.18$ & 6/121 & 15\\
       & 2760  & $\pi^-$ & $81\pm5/88\pm4$ & $0.21\pm0.01/0.26\pm0.01$ & $1.02\pm0.01/1.05\pm0.01$ & $0.49\pm0.02/0.86\pm0.05$ & $3.96\pm0.20/7.49\pm0.37$ & $16.56\pm0.80/17.47\pm0.87$ & $4.46\pm0.22/4.24\pm0.21$ & 15/114& 15\\
       &       & $\pi^+$ & $81\pm5/88\pm4$ & $0.21\pm0.01/0.26\pm0.01$ & $1.02\pm0.01/1.05\pm0.01$ & $0.49\pm0.03/0.86\pm0.05$ & $3.92\pm0.20/7.49\pm0.37$ & $16.55\pm0.80/17.47\pm0.87$ & $4.56\pm0.23/4.44\pm0.22$ & 15/115& 15\\
       & 7000  & $\pi^-$ & $83\pm4/89\pm4$ & $0.21\pm0.01/0.27\pm0.01$ & $1.02\pm0.01/1.05\pm0.01$ & $0.47\pm0.02/0.84\pm0.05$ & $3.98\pm0.19/7.69\pm0.38$ & $16.30\pm0.81/17.37\pm0.87$ & $5.55\pm0.28/5.41\pm0.27$ & 23/129& 15\\
       &       & $\pi^+$ & $82\pm4/87\pm4$ & $0.21\pm0.01/0.26\pm0.01$ & $1.02\pm0.01/1.05\pm0.01$ & $0.47\pm0.03/0.84\pm0.05$ & $3.99\pm0.19/7.59\pm0.38$ & $16.37\pm0.81/17.37\pm0.87$ & $5.60\pm0.28/5.60\pm0.28$ & 31/171& 15\\
       & 13000 & $\pi^-$ & $83\pm5/88\pm4$ & $0.22\pm0.01/0.27\pm0.01$ & $1.02\pm0.01/1.05\pm0.01$ & $0.46\pm0.02/0.82\pm0.05$ & $4.05\pm0.20/7.89\pm0.39$ & $16.31\pm0.82/17.27\pm0.86$ & $5.20\pm0.26/5.10\pm0.26$ & 13/50 & 15\\
       &       & $\pi^+$ & $84\pm5/86\pm4$ & $0.23\pm0.01/0.27\pm0.01$ & $1.02\pm0.01/1.05\pm0.01$ & $0.47\pm0.03/0.80\pm0.05$ & $4.09\pm0.20/7.99\pm0.40$ & $16.28\pm0.82/17.07\pm0.85$ & $5.20\pm0.26/5.10\pm0.26$ & 27/74 & 15\\
\hline
\end{tabular}}
\end{center}
\end{sidewaystable}
\begin{multicols}{2}

It should be noted that, from the fit process we know that, the
Tsallis expression, Eq. 2, having a polynomial behavior at large
$p_T$ could be a better description for the spectra at large
values of $p_T$, where the Boltzmann expression, Eq. (1) does not
work possibly at large $p_T$ if we use the same parameters $T_0$
and $\beta_T$. In some cases, for the spectra in a wide $p_T$
range, a single Tsallis expression is suitable, and two- or
three-component Boltzmann expression is needed. Our previous
work~\cite{42b} studied both the Tsallis and Boltzmann
distributions without flow effect and also confirms this issue.
Indeed, the Tsallis description is better than the Boltzmann one,
though the former has one more parameter $q$. In fact, the
introduction of the entropy index $q$ in the Tsallis description
has the meaning of reality. As discussed in section 2, $q$
characterizes the degree of non-equilibrium. In addition, when $q
\to 1$, the Tsallis description degenerates to the Boltzmann one.

To see clearly the excitation functions of free parameters,
Figures 2(a)--2(e) show the dependences of $T_0$, $\beta_T$,
$p_0$, $n$, and $k$ on $\sqrt{s}$, respectively. The blue and
black closed and open symbols represent the parameter values
corresponding to $\pi^-$ and $\pi^+$ respectively, which are
listed in Tables 1 and 2. The blue circles (black squares)
represent the first set of parameter values obtained from Eq. (7)
through Eqs. (1) [Eqs. (2)] and (3). The blue asterisks (black
triangles) represent the second set of parameter values obtained
from Eq. (7) through Eqs. (1) [Eqs. (2)] and (3). One can see that
the difference between the results of $\pi^-$ and $\pi^+$ is not
obvious. In the excitation functions of the first set of $T_0$ and
$\beta_T$ obtained from the blast-wave fit with Boltzmann
distribution, there are a hill at $\sqrt{s}\approx10$ GeV, a drop
at dozens of GeV, and then an increase from dozens of GeV to above
10 TeV. In the excitation functions of the first set of $T_0$ and
$\beta_T$ obtained from the blast-wave fit with Tsallis
distribution, there is no the complex structure, but a very low
hill. In the excitation functions of the second set of $T_0$ and
$\beta_T$, there is a slight increase from about 10 GeV to above
10 TeV. In Eq. (7) contained the blast-wave fit with both
distributions, in the excitation functions of $p_0$ and $n$, there
are a slight decrease and increase respectively in the case of the
hard component being available. The excitation function of $k$
shows that the contribution ($1-k$) of hard component slightly
increases from dozens of GeV to above 10 TeV, and it has no
contribution at around 10 GeV. At given energies, the fluctuations
in a given parameter result in different excitation functions due
to different selections. As a comparison, the red asterisks (green
triangles) in Fig. 2 represent the results from $AA$ collisions
which are discussed in detail in the appendix. One can see that
the results from $AA$ collisions approach to those from $pp$
collisions with the second set of parameters, though only the soft
component is used in most cases.

\begin{figure*}[!htb]
\begin{center}
\includegraphics[width=16.0cm]{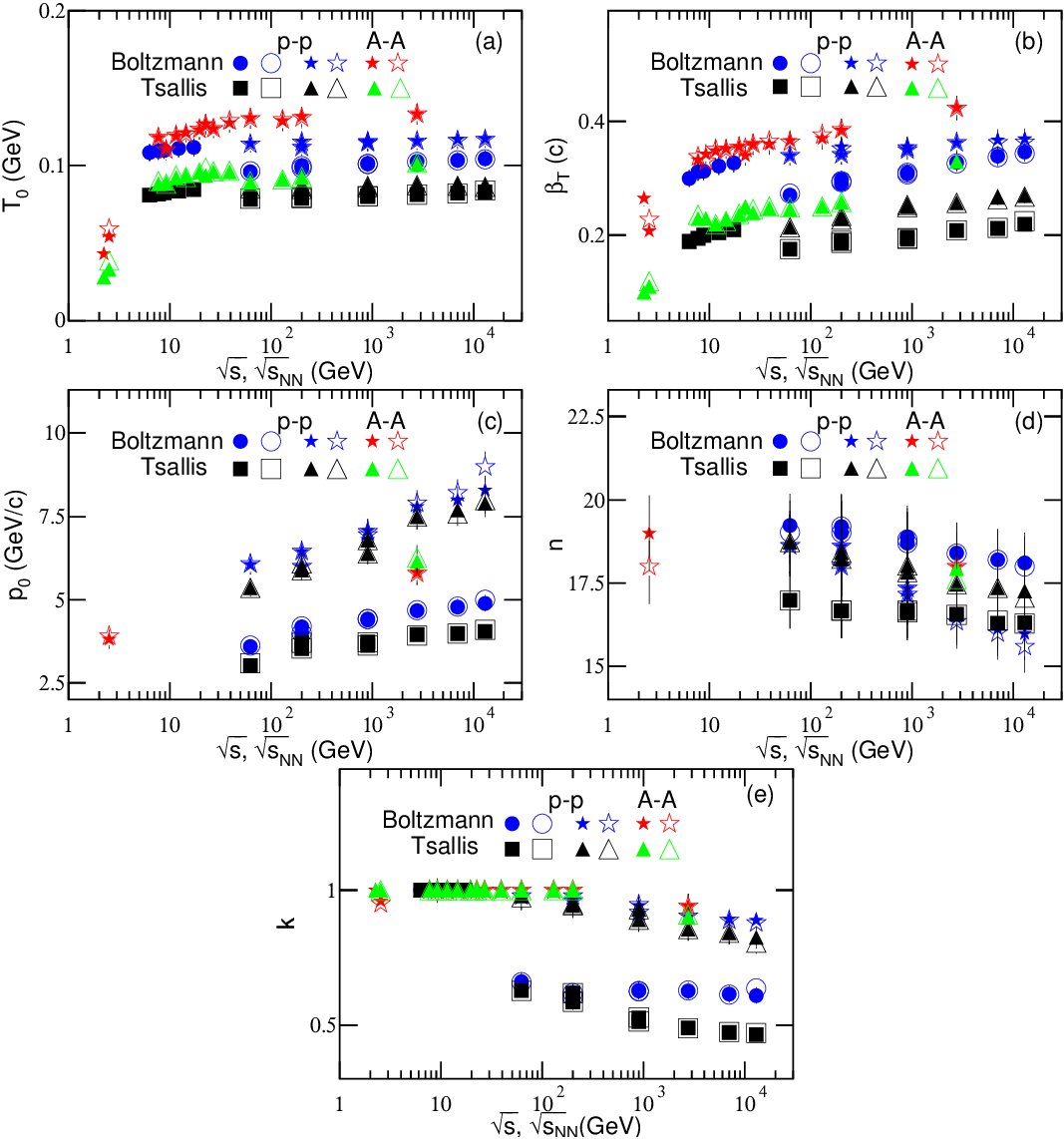}
\end{center}
{\small Fig. 2. Excitation functions of (a) $T_0$, (b) $\beta_T$,
(c) $p_0$, (d) $n$, and (e) $k$. The closed (open) symbols
represent the parameter values corresponding to $\pi^-$ ($\pi^+$)
spectra, which are listed in Tables 1 and 2. The blue circles
(black squares) represent the first set of parameter values
obtained from Eq. (7) through Eqs. (1) [Eqs. (2)] and (3). The
blue asterisks (black triangles) represent the second set of
parameter values obtained from Eq. (7) through Eqs. (1) [Eqs. (2)]
and (3). The red asterisks (green triangles) represent the
parameter values from $AA$ collisions for comparisons, which are
listed in Tables A1 and A2 in the appendix.}
\end{figure*}

\begin{figure*}[!htb]
\begin{center}
\includegraphics[width=16.0cm]{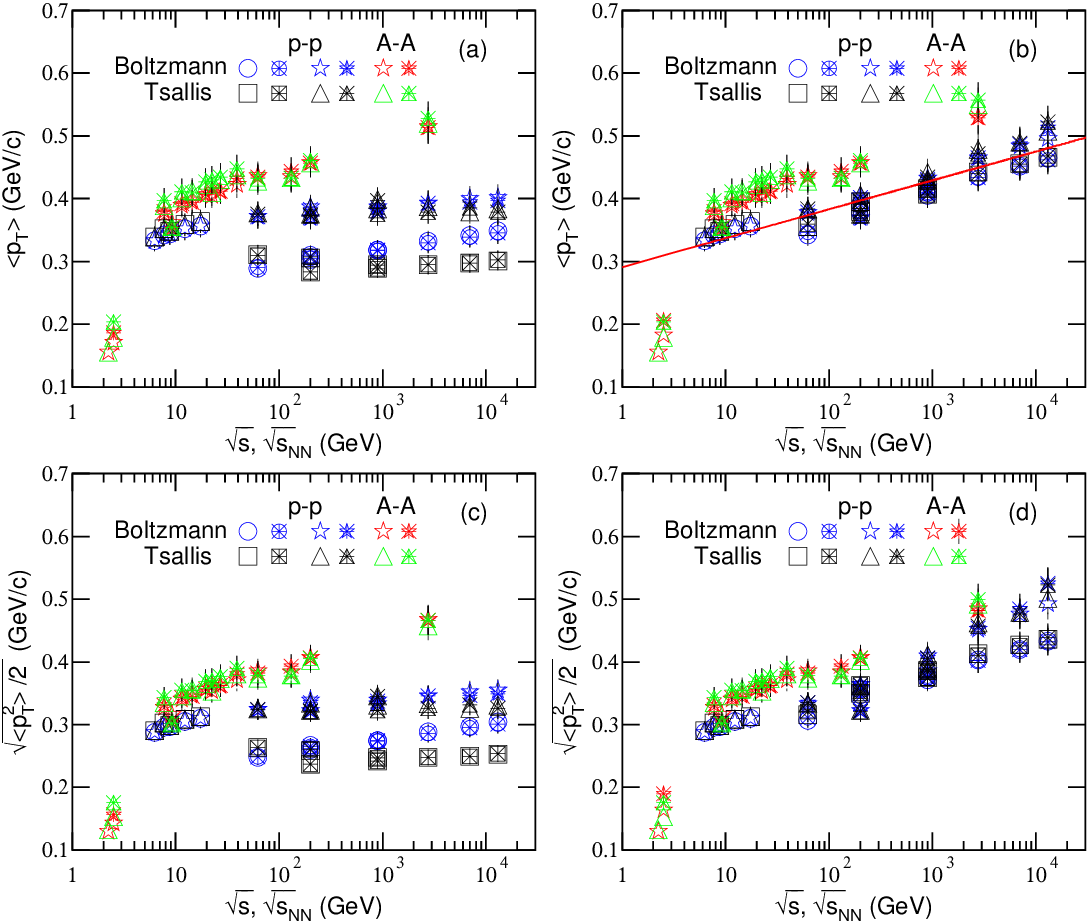}
\end{center}
{\small Fig. 3. Excitation functions of (a)(b) $\langle
p_T\rangle$ and (c)(d) $\sqrt{\langle p_T^2 \rangle/2}$. The open
symbols (open symbols with asterisks) represent the values
corresponding to $\pi^-$ ($\pi^+$) spectra. The blue circles
(black squares) represent the results obtained indirectly from Eq.
(1) [Eq. (2)] for the left panel, or from Eq. (7) through Eqs. (1)
[Eqs. (2)] and (3) for the right panel, by the first set of
parameter values. The results by the second set of parameter
values are presented by the blue asterisks (black triangles).
These values are indirectly obtained according to the parameter
values listed in Tables 1 and 2. The lines are the fitted results
for various symbols for $pp$ collisions. The red asterisks (green
triangles) represent the results from $AA$ collisions for
comparisons, which are indirectly obtained according to the
parameter values listed in Tables A1 and A2 in the appendix.}
\end{figure*}

\begin{figure*}[!htb]
\begin{center}
\includegraphics[width=16.0cm]{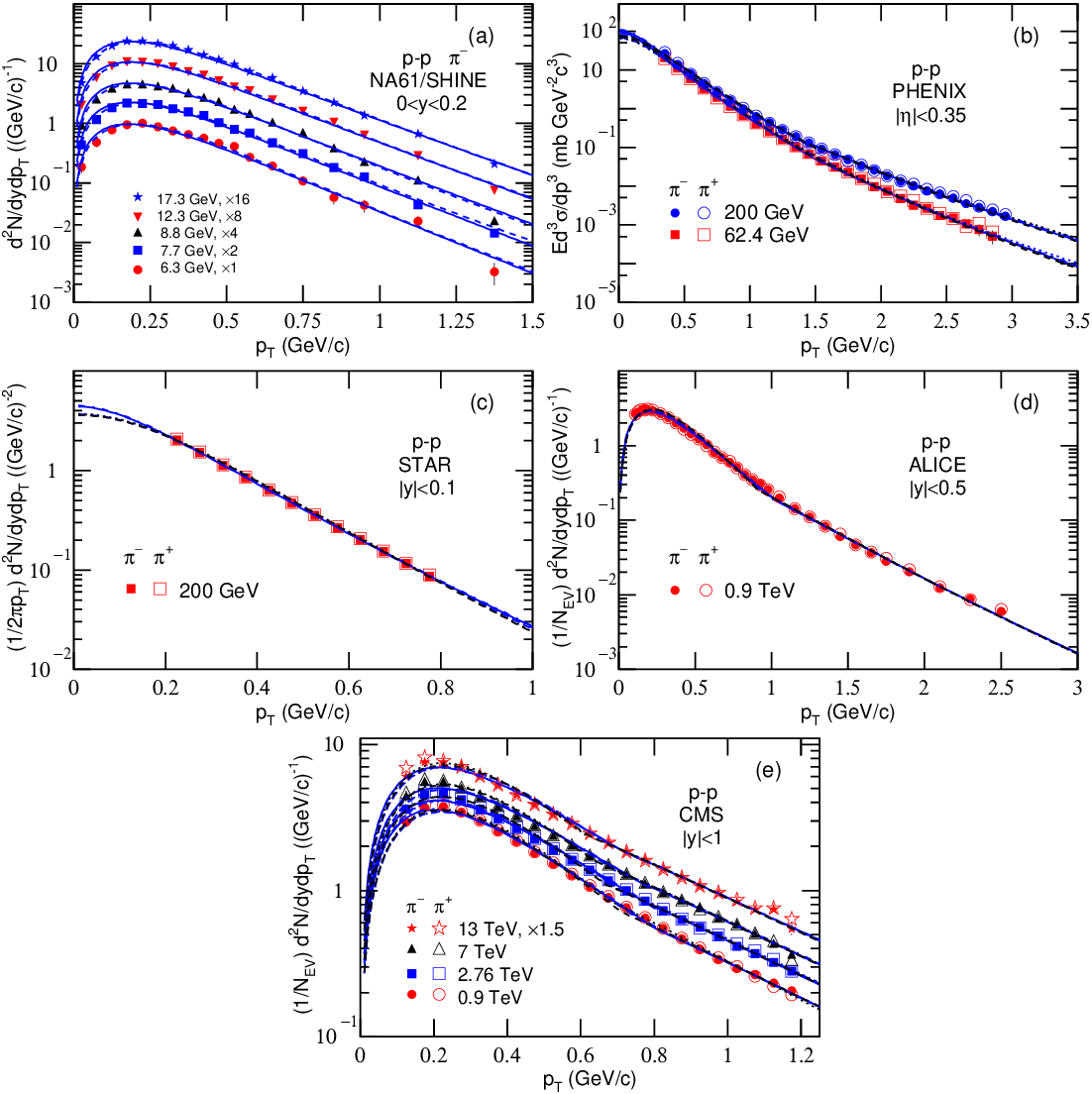}
\end{center}
{\small Fig. 4. Same as Fig. 1, but showing the results fitted by
Eq. (8) through Eqs. (1) and (3) with one set of parameter values
and by Eq. (8) through Eqs. (2) and (3) with two sets of parameter
values. The blue solid (dotted) curves are the results for $\pi^-$
($\pi^+$) spectra fitted by Eq. (8) through Eqs. (1) and (3), and
the blue and black dashed (dot-dashed) curves are the results for
$\pi^-$ ($\pi^+$) spectra fitted by Eq. (8) through Eqs. (2) and
(3).}
\end{figure*}

Indeed, $\sqrt{s_{NN}}\approx10$ GeV is a special energy for $AA$
collisions as indicated by Cleymans~\cite{43}. The present work
shows that $\sqrt{s}\approx10$ GeV is also a special energy for
$pp$ collisions. In particular, there is a hill in the excitation
functions of $T_0$ and $\beta_T$ in $pp$ collisions due to a given
selection of the parameters. At this energy (11 GeV more
specifically~\cite{43}), the final state has the highest net
baryon density, a transition from a baryon-dominated to a
meson-dominated final state takes place, and the ratios of strange
particles to mesons show clear and pronounced maximums~\cite{43}.
These properties result in this special energy.

At 11 GeV, the chemical freeze-out temperature in $AA$ collisions
is about 151 MeV~\cite{43}, and the present work shows that the
kinetic freeze-out temperature in $pp$ collisions is about 105
MeV, extracted from the blast-wave fit with Boltzmann
distribution. If we do not consider the difference between $AA$
and $pp$ collisions, though cold nuclear effect exists in $AA$
collisions, the chemical freeze-out happens obviously earlier than
the kinetic one. According to an ideal fluid consideration, the
time evolution of temperature follows
$T_f=T_i(\tau_i/\tau_f)^{1/3}$, where $T_i$ ($=300$ MeV) and
$\tau_i$ ($=1$ fm) are the initial temperature and proper time
respectively~\cite{44}\cite{45}, and $T_f$ and $\tau_f$ denote the
final temperature and time respectively, the chemical and kinetic
freeze-outs happen at 7.8 and 23.3 fm respectively. It should be
noted that in the calculation of $\tau_f$, the Lorentz factor is
not considered. If we consider the mean Lorentz factor
($\overline{\gamma}\approx5$--6) of charged pions in the rest
frame of emission source~\cite{15}--\cite{18}, the value of
freeze-out time will be smaller.

Strictly, $T_0$ ($\beta_T$) obtained from the pion spectra in the
present work is less than that averaged by weighting the yields of
pions, kaons, protons, and other light particles. Fortunately, the
fraction of the pion yield in high energy collisions are major
($\approx85\%$). The parameters and their tendencies obtained from
the pion spectra are similar to those obtained from the spectra of
all light particles. To study the excitation functions of $T_0$
and $\beta_T$, it does not matter if we use the spectra of pions
instead of all light particles.

It should be noted that the main parameters $T_0$ and $\beta_T$
are correlated in some way. Although the excitation functions of
$T_0$ ($\beta_T$) which are acceptable in the fit process are not
sole, their tendencies are harmonious in most cases, in particular
in the energy range from the RHIC to LHC. Combining with our
previous work~\cite{18}, we could say that there is a slight
($\approx10\%$) increase in the excitation function of $T_0$ and
an obvious ($\approx35\%$) increase in the excitation function of
$\beta_T$ from the RHIC to LHC. At least, the excitation functions
of $T_0$ and $\beta_T$ do not decrease from the RHIC to LHC.

However, the excitation function of $T_0$ from low to high
energies is not always incremental or invariant, though the
excitation function of $\beta_T$ has the trend of increase in
general. For example, In ref.~\cite{4}, $T_0$ slowly decreases as
$\sqrt{s}$ increases from 23 GeV to 1.8 TeV, and $\beta_T$ slowly
increases with $\sqrt{s}$. In refs.~\cite{19}\cite{20}, $T_0$ has
no obvious change and $\beta_T$ has a slight ($\approx10\%$)
increase from the RHIC to LHC. In ref.~\cite{21}, $T_0$ has a
slight ($\approx9\%$) increase and $\beta_T$ has a large
($\approx65\%$) increase from the RHIC to LHC. In
ref.~\cite{22}\cite{23}, $T_0$ has a slight ($\approx5\%$)
decrease from the RHIC to LHC and $\beta_T$ increases by
$\approx20\%$ from 39 to 200 GeV. It is convinced that $\beta_T$
increases from the RHIC to LHC, though the situation of $T_0$ is
doubtful.

Although some works~\cite{46}--\cite{49} reported a decrease of
$T_0$ and an increase of $\beta_T$ from the RHIC to LHC, our
re-scans on their plots show a different situation of $T_0$. For
example, in ref.~\cite{46}, our re-scans show that $T_0$ has no
obvious change and $\beta_T$ has a slight ($\approx9\%$) increase
from the top RHIC to LHC, though there is an obvious hill or there
is an increase by $\approx30\%$ in $T_0$ in 5--40 GeV comparting
with that at the RHIC. Ref.~\cite{47} shows similar results to
ref.~\cite{46} with the almost invariant $T_0$ from the top RHIC
to LHC, an increase by $\approx28\%$ in $T_0$ in 7--40 GeV
comparing with that at the top RHIC, and an increase by
$\approx8\%$ in $\beta_T$ comparing with that at the top RHIC.
Refs.~\cite{48}\cite{49} shows similar result to
refs.~\cite{46}\cite{47} on $T_0$, though the excitation function
of $\beta_T$ is not available.

In some cases, the correlation between $T_0$ and $\beta_T$ are not
negative, though some works~\cite{46}\cite{47} show negative
correlation over a wide energy range. For a give $p_T$ spectrum,
it seems that a larger $T_0$ corresponds to a smaller $\beta_T$,
which shows a negative correlation. However, this negative
correlation is not sole case. In fact, a couple of suitable $T_0$
and $\beta_T$ can fit a given $p_T$ spectrum. A series of $p_T$
spectra at different energies possibly show a positive correlation
between $T_0$ and $\beta_T$, or independent of $T_0$ on $\beta_T$,
in a narrow energy range. Very recently, ref.~\cite{50} shows
approximately independent of $T_0$ on $\beta_T$ in $pp$ collisions
at $\sqrt{s}=7$ TeV, and negative correlations in $p$-Pb
collisions at $\sqrt{s_{NN}}=5.02$ TeV and in Pb-Pb collisions at
$\sqrt{s_{NN}}=2.76$ TeV, for different average charged-particle
multiplicity densities, i.e. for different centrality classes.
These results are partly in agreement with our results. It seems
that the correlation between $T_0$ and $\beta_T$ is an open
question at present, though some researchers think that there is a
negative correlation between $T_0$ and $\beta_T$. In our opinion,
the type of correlation between $T_0$ and $\beta_T$ depends on
three factors, that is the choices of fitted region in low and
medium $p_T$ range, fixed or changeable $n_0$, sensitivity of
$\beta_T$ on centrality. Indeed, more studies are needed in the
near future.

\end{multicols}
\begin{sidewaystable}
{\small Table 3. Values of $T_0$, $\beta_T$, $k$, $p_0$, $n$,
$N_0$, $\chi^2$, and ndof corresponding to the solid (dotted)
curves for $\pi^-$ ($\pi^+$) spectra in Fig. 4, where Eq. (8)
through Eqs. (1) and (3) is used.} \vspace{-4mm}
\begin{center}
{\scriptsize
\begin{tabular} {ccccccccccc}\\ \hline\hline Collab. & $\sqrt{s}$ (GeV) &
Part. & $T_0$ (MeV) & $\beta_T$ ($c$) & $k$ & $p_0$ (GeV/c) & $n$ & $N_0$ & $\chi^2$ & ndof \\
\hline
NA61/  & 6.3   & $\pi^-$ & $105\pm5$ & $0.31\pm0.02$ & 1             & $-$           & $-$            & $0.08\pm0.01$  & 24 & 15\\
SHINE  & 7.7   & $\pi^-$ & $106\pm5$ & $0.32\pm0.02$ & 1             & $-$           & $-$            & $0.10\pm0.01$  & 44 & 15\\
INEL   & 8.8   & $\pi^-$ & $107\pm5$ & $0.32\pm0.02$ & 1             & $-$           & $-$            & $0.10\pm0.01$  & 86 & 15\\
       & 12.3  & $\pi^-$ & $108\pm5$ & $0.33\pm0.02$ & 1             & $-$           & $-$            & $0.12\pm0.01$  & 78 & 15\\
       & 17.3  & $\pi^-$ & $109\pm5$ & $0.33\pm0.02$ & 1             & $-$           & $-$            & $0.13\pm0.01$  & 34 & 15\\
\hline
PHENIX & 62.4  & $\pi^-$ & $111\pm5$ & $0.35\pm0.02$ & $0.99\pm0.01$ & $3.58\pm0.18$ & $19.26\pm0.56$ & $19.48\pm0.97$ & 9  & 20\\
INEL   &       & $\pi^+$ & $111\pm5$ & $0.35\pm0.02$ & $0.99\pm0.01$ & $3.59\pm0.18$ & $19.26\pm0.56$ & $19.54\pm0.97$ & 18 & 20\\
       & 200   & $\pi^-$ & $115\pm6$ & $0.37\pm0.02$ & $0.99\pm0.02$ & $4.20\pm0.21$ & $18.71\pm0.54$ & $24.11\pm1.20$ & 16 & 21\\
       &       & $\pi^+$ & $115\pm6$ & $0.36\pm0.02$ & $0.99\pm0.02$ & $4.31\pm0.22$ & $18.61\pm0.53$ & $24.96\pm1.22$ & 26 & 21\\
\hline
STAR   & 200   & $\pi^-$ & $114\pm6$ & $0.34\pm0.02$ & 1             & $-$           & $-$            & $0.26\pm0.01$  & 2  & 6\\
NSD    &       & $\pi^+$ & $114\pm6$ & $0.34\pm0.02$ & 1             & $-$           & $-$            & $0.27\pm0.01$  & 6  & 6\\
\hline
ALICE  & 900   & $\pi^-$ & $118\pm5$ & $0.35\pm0.02$ & $0.95\pm0.02$ & $4.41\pm0.22$ & $18.67\pm0.53$ & $3.70\pm0.18$  & 101& 27\\
INEL   &       & $\pi^+$ & $118\pm5$ & $0.35\pm0.02$ & $0.95\pm0.02$ & $4.40\pm0.22$ & $18.67\pm0.53$ & $3.69\pm0.18$  & 131& 27\\
\hline
CMS    & 900   & $\pi^-$ & $118\pm6$ & $0.35\pm0.02$ & $0.91\pm0.02$ & $4.03\pm0.20$ & $18.87\pm0.54$ & $8.90\pm0.44$  & 47 & 16\\
INEL   &       & $\pi^+$ & $118\pm5$ & $0.35\pm0.02$ & $0.91\pm0.02$ & $4.00\pm0.20$ & $18.67\pm0.55$ & $9.03\pm0.45$  & 43 & 16\\
       & 2760  & $\pi^-$ & $122\pm6$ & $0.36\pm0.02$ & $0.89\pm0.02$ & $4.01\pm0.18$ & $18.80\pm0.54$ & $11.34\pm0.57$ & 57 & 16\\
       &       & $\pi^+$ & $122\pm6$ & $0.37\pm0.02$ & $0.89\pm0.02$ & $4.02\pm0.18$ & $18.57\pm0.53$ & $11.54\pm0.58$ & 76 & 16\\
       & 7000  & $\pi^-$ & $123\pm6$ & $0.38\pm0.02$ & $0.87\pm0.02$ & $4.03\pm0.18$ & $18.50\pm0.52$ & $14.50\pm0.73$ & 73 & 16\\
       &       & $\pi^+$ & $123\pm6$ & $0.38\pm0.02$ & $0.86\pm0.02$ & $4.03\pm0.18$ & $18.40\pm0.52$ & $14.66\pm0.73$ & 90 & 16\\
       & 13000 & $\pi^-$ & $126\pm6$ & $0.37\pm0.02$ & $0.83\pm0.02$ & $4.04\pm0.19$ & $18.30\pm0.51$ & $13.62\pm0.68$ & 31 & 16\\
       &       & $\pi^+$ & $126\pm6$ & $0.37\pm0.02$ & $0.83\pm0.02$ & $4.04\pm0.19$ & $18.30\pm0.51$ & $13.84\pm0.69$ & 57 & 16\\
\hline
\end{tabular}}
\end{center}

{\small Table 4. Values of $T_0$, $\beta_T$, $q$, $k$, $p_0$, $n$,
$N_0$, $\chi^2$, and ndof corresponding to the dashed (dot-dashed)
curves for $\pi^-$ ($\pi^+$) spectra in Fig. 4, where Eq. (8)
through Eqs. (2) and (3) is used.} \vspace{-4mm}
\begin{center}
{\tiny
\renewcommand\tabcolsep{3.5pt}
\begin{tabular} {cccccccccccc}\\ \hline\hline Collab. & $\sqrt{s}$ (GeV) &
Part. & $T_0$ (MeV) & $\beta_T$ ($c$)& $q$ & $k$ & $p_0$ (GeV/c) & $n$ & $N_0$ & $\chi^2$ & ndof \\
\hline
NA61/  & 6.3   & $\pi^-$ & $83\pm5$ & $0.24\pm0.01$ & $1.04\pm0.01$ & 1 & $-$ & $-$ & $0.09\pm0.01$ & 11 & 14\\
SHINE  & 7.7   & $\pi^-$ & $84\pm5$ & $0.25\pm0.02$ & $1.04\pm0.01$ & 1 & $-$ & $-$ & $0.10\pm0.01$ & 8  & 14\\
INEL   & 8.8   & $\pi^-$ & $84\pm5$ & $0.25\pm0.01$ & $1.04\pm0.01$ & 1 & $-$ & $-$ & $0.10\pm0.01$ & 23 & 14\\
       & 12.3  & $\pi^-$ & $85\pm5$ & $0.26\pm0.01$ & $1.05\pm0.01$ & 1 & $-$ & $-$ & $0.12\pm0.01$ & 11 & 14\\
       & 17.3  & $\pi^-$ & $86\pm5$ & $0.26\pm0.01$ & $1.05\pm0.01$ & 1 & $-$ & $-$ & $0.13\pm0.01$ & 4  & 14\\
\hline
PHENIX & 62.4  & $\pi^-$ & $82\pm4/88\pm4$ & $0.24\pm0.01/0.27\pm0.01$ & $1.07\pm0.01/1.05\pm0.01$ & $0.99\pm0.01/0.99\pm0.01$ & $3.20\pm0.19/3.19\pm0.18$ & $18.56\pm0.51/18.56\pm0.51$ & $18.27\pm0.91/19.18\pm0.93$ & 33/39 & 18\\
INEL   &       & $\pi^+$ & $82\pm5/88\pm4$ & $0.24\pm0.01/0.27\pm0.01$ & $1.07\pm0.01/1.06\pm0.01$ & $0.99\pm0.01/0.99\pm0.01$ & $3.20\pm0.19/3.21\pm0.18$ & $18.56\pm0.51/18.56\pm0.51$ & $17.19\pm0.89/17.36\pm0.90$ & 15/15 & 18\\
       & 200   & $\pi^-$ & $83\pm5/90\pm4$ & $0.25\pm0.02/0.28\pm0.01$ & $1.07\pm0.01/1.06\pm0.01$ & $0.99\pm0.01/0.99\pm0.01$ & $3.99\pm0.19/3.89\pm0.19$ & $18.06\pm0.51/18.06\pm0.51$ & $23.21\pm1.15/22.40\pm1.10$ & 16/31 & 19\\
       &       & $\pi^+$ & $83\pm5/90\pm4$ & $0.25\pm0.02/0.28\pm0.01$ & $1.07\pm0.01/1.06\pm0.01$ & $0.99\pm0.01/0.99\pm0.01$ & $4.09\pm0.20/4.09\pm0.20$ & $18.01\pm0.50/18.01\pm0.51$ & $23.00\pm1.16/22.40\pm1.14$ & 26/32 & 19\\
\hline
STAR   & 200   & $\pi^-$ & $83\pm5/89\pm4$ & $0.25\pm0.01/0.28\pm0.01$ & $1.06\pm0.01/1.04\pm0.01$ & 1 & $-$ & $-$ & $0.26\pm0.01$ & 34 & 5\\
NSD    &       & $\pi^+$ & $83\pm5/89\pm4$ & $0.25\pm0.01/0.28\pm0.01$ & $1.06\pm0.01/1.04\pm0.01$ & 1 & $-$ & $-$ & $0.26\pm0.01$ & 38 & 5\\
\hline
ALICE  & 900   & $\pi^-$ & $84\pm5/91\pm4$ & $0.26\pm0.01/0.30\pm0.01$ & $1.06\pm0.01/1.03\pm0.01$ & $0.94\pm0.02/0.94\pm0.02$ & $4.11\pm0.19/4.11\pm0.19$ & $17.99\pm0.50/17.99\pm0.50$ & $0.59\pm0.02/0.58\pm0.02$ & 187/287 & 25\\
INEL   &       & $\pi^+$ & $85\pm4/91\pm4$ & $0.26\pm0.01/0.30\pm0.01$ & $1.06\pm0.01/1.03\pm0.01$ & $0.94\pm0.02/0.94\pm0.02$ & $4.11\pm0.20/4.11\pm0.20$ & $17.99\pm0.40/17.99\pm0.40$ & $0.59\pm0.02/0.58\pm0.02$ & 232/333 & 25\\
\hline
CMS    & 900   & $\pi^-$ & $87\pm4/92\pm4$ & $0.27\pm0.01/0.30\pm0.01$ & $1.05\pm0.01/1.03\pm0.01$ & $0.88\pm0.02/0.87\pm0.02$ & $3.85\pm0.19/3.85\pm0.19$ & $18.40\pm0.51/18.40\pm0.51$ & $1.43\pm0.26/1.42\pm0.27$ & 79/95 & 14\\
INEL   &       & $\pi^+$ & $87\pm4/93\pm4$ & $0.27\pm0.01/0.30\pm0.01$ & $1.05\pm0.01/1.03\pm0.01$ & $0.88\pm0.02/0.86\pm0.02$ & $3.82\pm0.19/3.82\pm0.19$ & $18.53\pm0.51/18.53\pm0.51$ & $1.42\pm0.25/1.42\pm0.24$ & 70/72 & 14\\
       & 2760  & $\pi^-$ & $91\pm5/93\pm5$ & $0.30\pm0.01/0.30\pm0.02$ & $1.04\pm0.01/1.03\pm0.01$ & $0.83\pm0.02/0.83\pm0.02$ & $3.92\pm0.20/3.92\pm0.20$ & $18.43\pm0.52/18.43\pm0.52$ & $1.67\pm0.27/1.81\pm0.27$ & 70/72 & 14\\
       &       & $\pi^+$ & $92\pm5/93\pm5$ & $0.30\pm0.02/0.30\pm0.02$ & $1.04\pm0.01/1.03\pm0.01$ & $0.84\pm0.02/0.84\pm0.02$ & $3.92\pm0.20/3.92\pm0.20$ & $18.43\pm0.51/18.43\pm0.51$ & $1.83\pm0.27/1.82\pm0.26$ & 87/88 & 14\\
       & 7000  & $\pi^-$ & $92\pm4/94\pm5$ & $0.30\pm0.02/0.31\pm0.02$ & $1.04\pm0.01/1.03\pm0.01$ & $0.79\pm0.02/0.81\pm0.02$ & $3.94\pm0.19/3.94\pm0.19$ & $18.41\pm0.50/18.41\pm0.50$ & $2.30\pm0.32/2.30\pm0.33$ & 68/69 & 14\\
       &       & $\pi^+$ & $91\pm4/94\pm5$ & $0.30\pm0.01/0.31\pm0.02$ & $1.03\pm0.01/1.03\pm0.01$ & $0.79\pm0.02/0.81\pm0.02$ & $3.95\pm0.20/3.95\pm0.20$ & $18.41\pm0.50/18.41\pm0.50$ & $2.31\pm0.33/2.30\pm0.32$ & 87/92 & 14\\
       & 13000 & $\pi^-$ & $91\pm5/95\pm5$ & $0.30\pm0.01/0.31\pm0.02$ & $1.04\pm0.01/1.02\pm0.01$ & $0.81\pm0.02/0.80\pm0.02$ & $3.96\pm0.19/3.96\pm0.19$ & $18.31\pm0.41/18.31\pm0.41$ & $2.19\pm0.39/2.17\pm0.38$ & 34/39 & 14\\
       &       & $\pi^+$ & $91\pm5/95\pm5$ & $0.30\pm0.01/0.31\pm0.02$ & $1.03\pm0.01/1.02\pm0.01$ & $0.78\pm0.02/0.79\pm0.02$ & $3.96\pm0.19/3.96\pm0.19$ & $18.31\pm0.41/18.31\pm0.41$ & $2.18\pm0.38/2.17\pm0.36$ & 56/59 & 14\\
\hline
\end{tabular}}
\end{center}
\end{sidewaystable}
\begin{multicols}{2}

\begin{figure*}[!htb]
\begin{center}
\includegraphics[width=16.0cm]{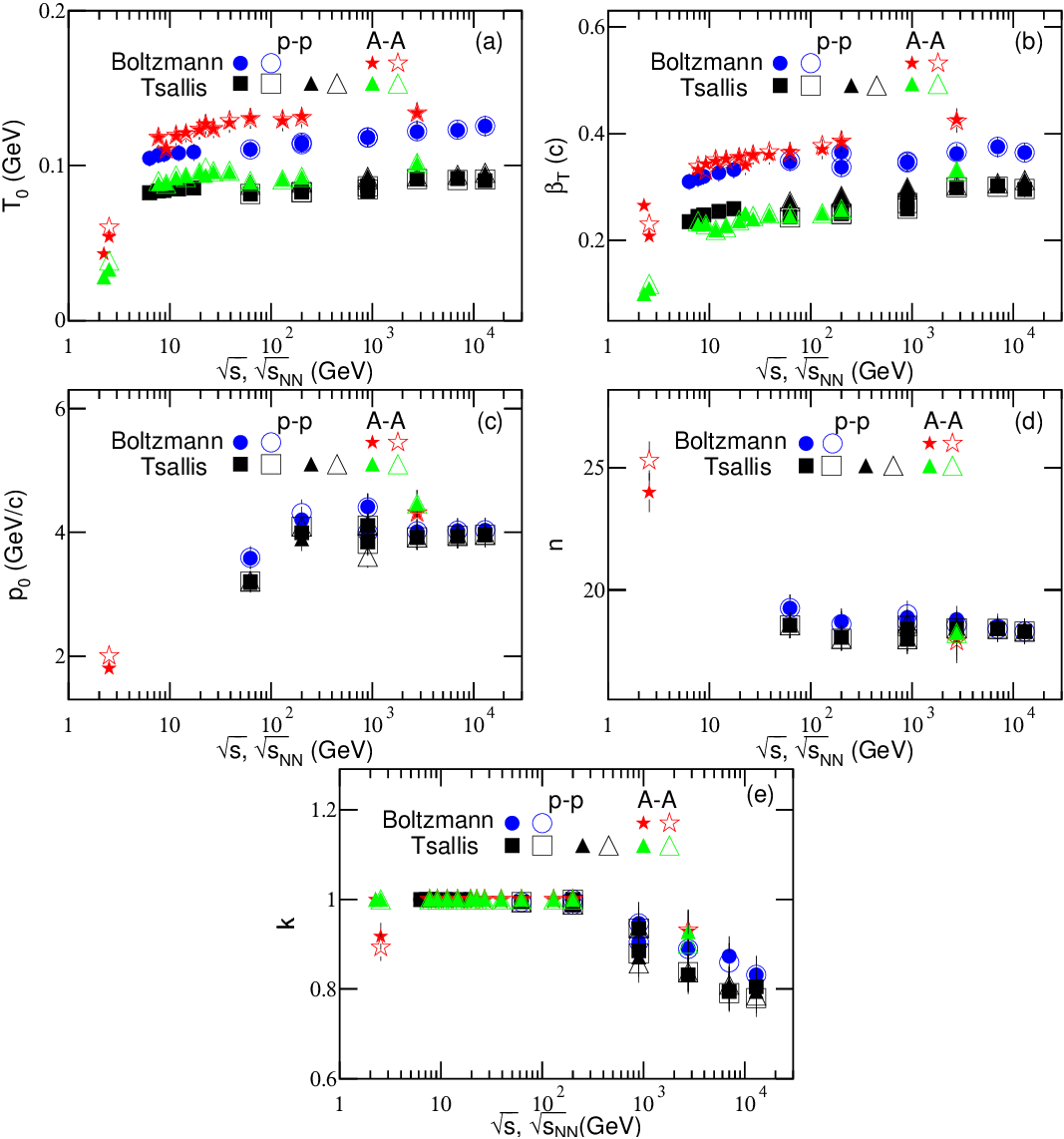}
\end{center}
{\small Fig. 5. Same as Fig. 2, but showing the results obtained
from Eq. (8) through Eqs. (1) and (3) with one set of parameter
values and from Eq. (8) through Eqs. (2) and (3) with two sets of
parameter values. The blue circles represent the parameter values
obtained from Eq. (8) through Eqs. (1) and (3). black The squares
(triangles) represent the first (second) set of parameter values
obtained from Eq. (8) through Eqs. (2) and (3). The related
parameter values are listed in Tables 3 and 4. The red asterisks
(green triangles) represent the parameter values from $AA$
collisions for comparisons, which are listed in Tables A3 and A4
in the appendix.}
\end{figure*}

\begin{figure*}[!htb]
\begin{center}
\includegraphics[width=16.0cm]{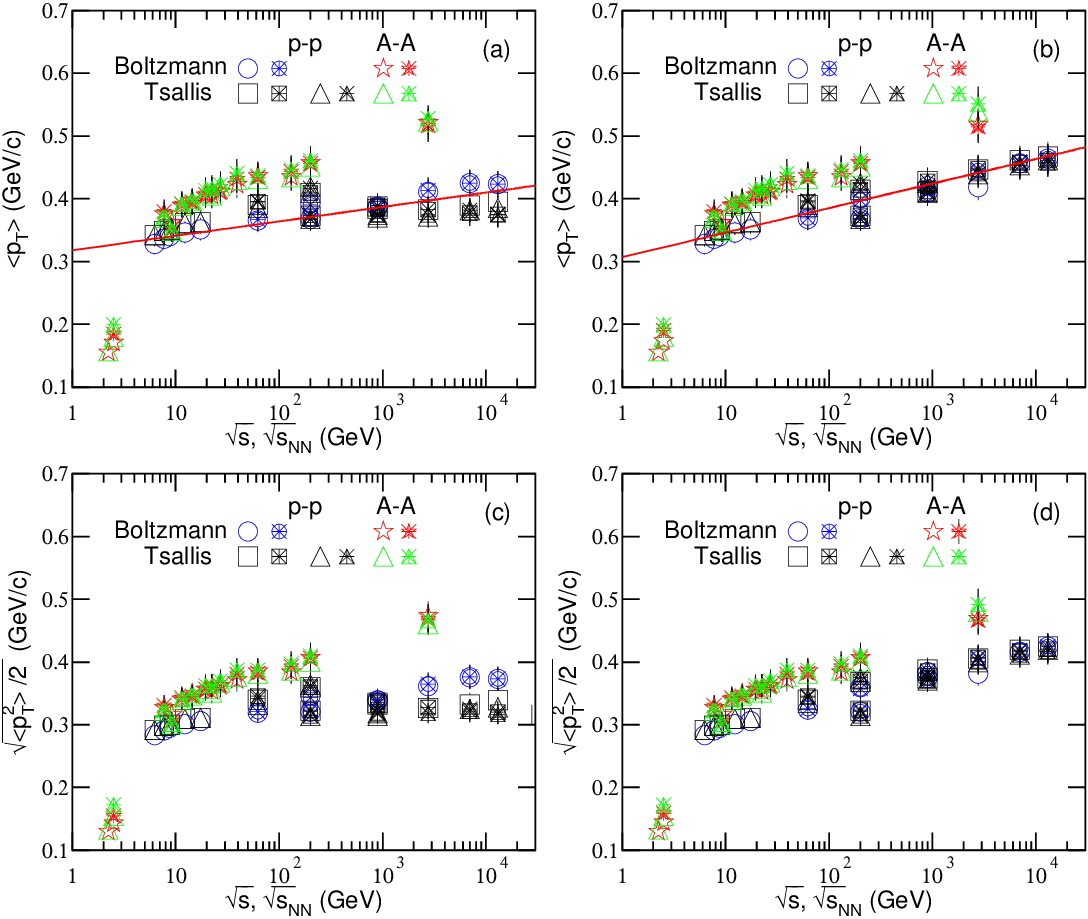}
\end{center}
{\small Fig. 6. Same as Fig. 3, but showing the results obtained
from Eq. (1) [Eq. (2)] for the left panel, or from Eq. (8) through
Eqs. (1) [Eqs. (2)] and (3) for the right panel. The circles
represent the results obtained indirectly from Eq. (1) for the
left panel, or from Eq. (8) through Eqs. (1) and (3) for the right
panel, by the parameter values. The squares (triangles) represent
the results obtained indirectly from Eq. (2) for the left panel,
or from Eq. (8) through Eqs. (2) and (3) for the right panel, by
the first (second) set of parameter values. These values are
indirectly obtained according to the parameter values listed in
Tables 3 and 4. The red asterisks (green triangles) represent the
results from $AA$ collisions for comparisons, which are indirectly
obtained according to the parameter values listed in Tables A3 and
A4 in the appendix.}
\end{figure*}

To study further the behaviors of parameters, Figure 3 shows the
excitation functions of (a)(b) mean $p_T$ ($\langle p_T \rangle$)
and (c)(d) ratio of root-mean-square $p_T$ ($\sqrt{\langle p_T^2
\rangle}$) to $\sqrt{2}$, where the left panel [(a)(c)]
corresponds to the results of the first component [Eq. (1) or (2)]
and the right panel [(b)(d)] corresponds to the results of the two
components [Eq. (7)]. The open symbols (open symbols with
asterisks) represent the values corresponding to $\pi^-$ ($\pi^+$)
spectra. The blue circles (black squares) represent the values
obtained indirectly from Eq. (1) [Eq. (2)] for the left panel or
Eq. (7) through Eqs. (1) [Eqs. (2)] and (3) for the right panel,
by the first set of parameter values. The results by the second
set of parameter values are presented by the blue asterisks (black
triangles). These values are indirectly obtained from the
equations according to the parameters listed in Tables 1 and 2
over a $p_T$ range from 0 to 5 GeV/$c$ which is beyond the
available range of the data. If the initial temperature of
interacting system is approximately presented by
$T_i=\sqrt{\langle p_T^2 \rangle/2}$~\cite{51}--\cite{53}, the
lower panel shows the excitation function of initial temperature.
Because of excluding different contribution fractions of the
second component, the left panel [(a)(c)] shows some differences
in the case of using two sets of parameter values. It should be
noted that the root-mean square momentum component of particles in
the rest frame of isotropic emission source is regarded as the
initial temperature, or at the least it is a reflection of the
initial temperature. The relations in the left panel are complex
and multiple due to different sets of parameter values. The line
in Fig. 3(b) is fitted to various symbols by linear function
\begin{align}
\langle p_T \rangle=(0.291\pm0.006)+(0.020\pm0.001)\ln\sqrt{s}
\end{align}
with $\chi^2$/ndof=54/82. From the line one can see that the
behavior of $\langle p_T \rangle$. In particular, with the
increase of $\ln\sqrt{s}$ and including the contribution of second
component, $\langle p_T \rangle$ increases approximately linearly.
As a comparison, the red asterisks (green triangles) in Fig. 3
represent the results for $AA$ collisions, which are indirectly
obtained according to the parameter values listed in Tables A1 and
A2. One can see that the results for $AA$ collisions are greater
than those for $pp$ collisions at around 10 GeV and above.

The quantities $\langle p_T \rangle$ and $T_i$ are very important
to understand the excitation degree of interacting system. As for
the right panel in Fig. 3 which is for the two-component, the
incremental trend for $\langle p_T \rangle$ and $T_i$ with the
increase of $\sqrt{s}$ ($\sqrt{s_{NN}}$) is a natural result due
to more energy deposition at higher energy. Although $\langle p_T
\rangle$ and $T_i$ are obtained from the parameter values listed
in Tables 1 and 2 (A1 and A2), they are independent of fits or
models. More investigations on the excitation functions of
$\langle p_T \rangle$ and $T_i$ are needed due to their
importance.
\\

{\subsection {Comparison with data by Eq. (8)}}

To discuss further, for comparisons with the results from Eq. (7),
we reanalyze the spectra by Eq. (8) and study the trends of new
parameters. Figure 4 is the same as Fig. 1, but showing the
results fitted by Eq. (8) through Eqs. (1) and (3) and through
Eqs. (2) and (3) respectively. For Eq. (8) through Eqs. (1) and
(3), only one set of parameter values are used due to the fact
that there is no correlation in the extraction of parameters in
the two-component fit. For Eq. (8) through Eqs. (2) and (3), two
sets of parameter values are used to see the insensitivity of main
parameters. The first set of parameter values are obtained by the
method of least squares. The second set of parameter values are
obtained by increasing or decreasing main parameters ($T_0$,
$\beta_T$, $q$, and $k$) by a few percent, and limits the increase
of $\chi^2$ by a few percent. The values of related parameters are
listed in Tables 3 and 4 which are the same as Tables 1 and 2
respectively, and with only one set of parameter values in Table
3. The related parameters are shown in Fig. 5 and the leading-out
parameters are shown in Fig. 6, which are the same as Figs. 2 and
3 respectively, and with only one set of parameter values for Eq.
(8) through Eqs. (1) and (3). In particular, $k$ in Fig. 5 is
obtained by $k=\int_0^{p_1}A_1f_S(p_T)dp_T$ due to $f_0(p_T)$ is
normalized to 1, as discussed following Eq. (8). The lines in
Figs. 6(a) and 6(b) are fitted by linear functions
\begin{align}
\langle p_T \rangle=(0.318\pm0.004)+(0.010\pm0.001)\ln\sqrt{s}
\end{align}
and
\begin{align}
\langle p_T \rangle=(0.307\pm0.005)+(0.017\pm0.001)\ln\sqrt{s}
\end{align}
with $\chi^2$/ndof=55/61 and 28/61 respectively, though the linear
relationships between the parameters and $\ln\sqrt{s}$ may be not
the best fitting functions. Similar to Figs. 2 and 3, the results
for $AA$ collisions from Eq. (8) are also presented in Figs. 5 and
6 for comparisons. One can see the similarity in both $pp$ and
$AA$ collisions in the considered energy range.

From Fig. 4 one can see that Eq. (8) fits similarly good the data
as Eq. (7). Because of the data being not available in very-low
$p_T$ range, Eq. (10) is not used in the fit. $T_0$ and $\beta_T$
in Fig. 5 increase slightly from a few GeV to above 10 TeV with
some fluctuations in some cases, which is partly similar to those
in Fig. 2. Other parameters in Fig. 5 show somehow similar trends
to Fig. 2 with some differences. The left panels in Figs. 6 and 3
are different due to the first component being in different
superpositions. The right panels in Figs. 6 and 3 are very similar
to each other due to the same data sets being fitted. We would
like to point out that Figs. 1 and 4 present various cases which
are displayed together and result in Figs. 2 and 3 as well as
Figs. 5 and 6 depicting much more data in which some energy points
are taken from Figs. A1 and A2 in the appendix.

Before continuing this work, we would like to point out the
justification and correctness for the comparisons of $pp$ and
central $AA$ collisions in Figs. 2, 3, 5, and 6. No matter
peripheral or central $AA$ collisions, a set of nucleon-nucleon
collisions in participant region are similar to the minimum-bias
$pp$ collisions. Peripheral $AA$ collisions are similar to central
collisions with smaller projectile and target nuclei, while
central $AA$ collisions are just central collisions with large
nuclei. Because of collision energies considered in this work are
high, the nucleon-nucleon correlation, cluster structure, medium
effect, and other nuclear effects in participant region can be
neglected. Meanwhile, the cold nuclear or spectator effect in
non-central $AA$ collisions can be neglected, too. In our opinion,
we may compare the minimum-bias $pp$ collisions with the $AA$
collisions in any centrality class.

Combining with our recent work~\cite{54}, it is shown the
similarity in $pp$ and $AA$ collisions, though $AA$ collisions
appear larger $T_0$, $\beta_T$, $\langle p_T\rangle$, and $T_i$ in
most cases. Moreover, it is well seen that Tsallis does not
distinguish well between the data in $pp$ and $AA$
collisions~\cite{55}--\cite{57}. Indeed, at high energy, both $pp$
and $AA$ collisions produce many particles and obey statistical
law. In addition, $pp$ collisions are the basic sub-process in
$AA$ collisions. It is natural that $pp$ and $AA$ collisions show
similar law. However, concerning around 10--20 GeV change, this is
expected as soon as QCD effects/calculations may not be directly
applicable below this point, plus seem to be smoothed away by
other processes in $AA$ collisions. This is well visible as a
clear difference as shown in refs.~\cite{55}--\cite{57}, where the
data in $pp$ collisions goes well with the data in
electron-positron collisions, while the data in $AA$ collisions
are different.

The differences between Eqs. (7) and (8) are obviously, though the
similar components are used in them. In our recent
works~\cite{18}\cite{58}, Eqs. (7) and (8) are used respectively.
Although there is correlation in the extraction of parameters, a
smooth curve can be easily obtained by Eq. (7). Although it is not
easy to obtain a smooth curve at the point of junction, there is
no or less correlation in the extraction of parameters by Eq. (8).
In consideration of obtaining a set of parameters with least
correlation, we are inclined to use Eq. (8) to extract the related
parameters. This inclining results in Eq. (8) to separate
determinedly the contributions of soft and hard processes.

It should be noted that the system of $pp$ collisions at low
energy is probably not in thermal equilibrium or local thermal
equilibriums due to low multiplicity, which is not the case of the
present work. In fact, the present work treats $pp$ collisions at
high energies in which the multiplicities in most cases are not
too low. In addition, related review work~\cite{59} shows that
small system also appears similar collective behavior to $AA$
collisions. This renders that the idea of local thermal
equilibrium is suitable to high energy $pp$ collisions for which
the blast-wave fit can be used, though QGP is not expected to form
in minimum-bias events.

Although we have used the blast-wave fit in the superposition
function with two components in which the second component is an
inverse power-law, the blast-wave part is not necessary for
fitting process itself. In fact, the superposition of
(two-)Boltzmann (or Tsallis) distribution and inverse power-law
can fit the data in most cases~\cite{29}\cite{30}\cite{57}. In
particular, in our very recent work~\cite{60}, we used the
Tsallis--Pareto-type function~\cite{28}\cite{61}\cite{62} to fit
$p_T$ spectra in a wide range, in which there is no boosted part.
The merits of the boosted part are that some additional quantities
such as $T_0$ and $\beta_T$ can be obtained and physics picture is
more abundant.

To avoid the dependences of $T_0$ and $\beta_T$ on fits or models,
one can use $\langle p_T\rangle$ to describe synchronously $T_0$
and $\beta_T$. Generally, $\langle p_T\rangle$ is independent of
fits or models, though it can be calculated from fits or models.
Averagely, the contribution of one participant in each binary
collisions is $\langle p_T \rangle/2$ which is resulted from both
the thermal motion and flow effect. Let $k_0$ denote the
contribution fraction of thermal motion. The contribution fraction
of flow effect is naturally $1-k_0$. Thus, we define
$T_0=k_0\langle p_T \rangle/2$ and $\beta_T=(1-k_0)\langle p_T
\rangle/2m_0\overline{\gamma}$. As a free parameter, $k_0$ depends
on collision energy, which is needed to study further.
\\

{\subsection {More discussions}}

Before summary and conclusions, we would like to underline the
preponderance of the present work. Comparing with PYTHIA or other
perturbative QCD simulation tools~\cite{63}--\cite{66}, the
present work is simpler and more applicative in obtaining the
excitation functions of $T_0$ and $\beta_T$, though the usual
fitting method is used. In a recent work~\cite{67}, it was pointed
out that the PYTHIA Monte Carlo~\cite{63}\cite{64} disagrees with
some data, and the two-component (soft+hard) model describes the
data accurately and comprehensively, though the two-component
model used in ref.~\cite{67} is different from the fit used in the
present work. In addition, the present work is a data-driven
reanalysis based on some physics considerations, but not a simple
fit to the data. From the data-driven reanalysis, the excitation
functions of some quantities have been obtained.

From the excitation functions, one can see some complex structures
which are useful to study the properties of particle production
and system evolution at different energies. In particular, in the
energy range around 10 GeV, the excitation functions have
transition which implies the phase of interaction matter had
changed. In addition, by using the two-component fit, the present
work also presents a new method to extract the contribution
fraction of hard component. One can see that this contribution
fraction is 0 in the energy range around 10 GeV. This implies that
the interactions in the energy range around 10 GeV have only soft
component, and that above 10 GeV have both soft and hard
components. The interaction mechanisms in the two energy ranges
are different.

We would like to emphasize that the aim of the present work is to
look for possible signatures of a transition from baryon-dominated
to meson-dominated hadron production mechanism by fitting pion
$p_T$ spectra at mid-(pseudo)rapidity in $pp$ collisions which
also show collective phenomenon~\cite{59}. The main conclusion
regards a possible indication of such an effect at about 10--20
GeV visible in a drop of the temperature extracted using a
blast-wave model with Boltzmann distribution to model the soft
excitation mode. Such a conclusion seems not straightforward since
the results are strongly biased by different $p_{T\max}$ used to
fit data at different energies. In fact, the results are less
affected by $p_{T\max}$ due to the fact that the parameters are
mainly determined by the spectra in low and intermediate $p_T$
regions. Larger $p_{T\max}$ does not change the trend of inverse
power-law, which does not affect obviously the parameters.
Although $p_{T\max}$ has no obvious influence on the parameters,
we have used $p_{T\max}=5$ Gev/$c$ in our calculation.

It should be noted that the blast-wave analysis is known to be
sensitive to the selected $p_T$ range in the case of using a not
too wide and local one such as 1--2 GeV/$c$. To avoid this
dependence, we have used a wide enough $p_T$ range from 0 to a
large value, but not a narrow and local one. In addition, looking
at the fit results when using a Tsallis function to describe soft
excitation processes such a drop in the $k$-parameter is no longer
visible which seems to mean that the fit is less sensitive to the
hard component and also less sensitive to $p_{T\max}$. Using the
Tsallis function assumption also the drop in the temperature is no
longer visible which means that the effect on the temperature seen
with the Boltzmann assumption seems to be model dependent. In
fact, the parameters discussed in the present work are indeed
model dependent. We hope to extract model independent parameters
in the near future. Our definition $T_0=k_0\langle p_T \rangle/2$
and $\beta_T=(1-k_0)\langle p_T \rangle/2m_0\overline{\gamma}$ are
a possible choice for the model independent parameters.

On the other hand, although comparing Boltzmann with Tsallis is an
interesting physical problem, which could not be understood
without another comparison with the generic super-statistics, as
the one implemented in particle productions~\cite{68}--\cite{71}.
The latter -- in contrary to Boltzmann and Tsallis -- lets the
system alone judge about its statistical nature, whether extensive
or non-extensive (equilibrium or non-equilibrium). As pointed out
in refs.~\cite{68}--\cite{71}, the particle production at BES
energies is likely a non-extensive process but not necessarily
Boltzmann or Tsallis type. Indeed, further study on particle
production in high energy collisions is needed in the future.

In particular, larger $T_0$ means higher excitation degree and
shorter lifetime of the fireball formed in high energy collisions.
At the LHC energy, the fireball should have higher excitation
degree comparing with that at the top RHIC energy, though longer
lifetime is possible at the LHC energy~\cite{47}\cite{72}. As a
result of competition between excitation degree and lifetime,
$T_0$ shows the trend of increase with energy in the present work.
Meanwhile, our result on larger $\beta_T$ means quicker expansion
at the LHC energy. These trends are harmonious with those of
$\langle p_T\rangle$ and $T_i$ as well as other method such as the
alternative method~\cite{16}\cite{18}\cite{58} by which $T_0$ and
$\beta_T$ are also obtained.
\\

{\section{Summary and conclusion}}

The transverse momentum spectra of $\pi^-$ and $\pi^+$ produced at
mid-(pseudo)rapidity in $pp$ collisions over an energy range from
a few GeV to above 10 TeV have been analyzed by the superposition
of the blast-wave fit with Boltzmann distribution or with Tsallis
distribution and the inverse power-law (Hagedorn function). The
fit results are well fitting to the experimental data of
NA61/SHINE, PHENIX, STAR, ALICE, and CMS Collaborations. The
values of related parameters are extracted from the fit process
and the excitation functions of parameters are obtained.

In the particular superposition Eq. (7) and with a given
selection, both excitation functions of $T_0$ and $\beta_T$
obtained from the blast-wave fit with Boltzmann distribution show
a hill at $\sqrt{s}\approx10$ GeV, a drop at dozens of GeV, and an
increase from dozens of GeV to above 10 TeV. The mentioned two
excitation functions obtained from the blast-wave fit with Tsallis
distribution does not show such a complex structure, but a very
low hill. In another selection for the parameters in Eq. (7) or in
the superposition Eq. (8), $T_0$ and $\beta_T$ increase generally
quickly from a few GeV to about 10 GeV and then slightly at above
10 GeV. There is no the complex structure, too. In both
superpositions, the excitation function of $p_0$ ($n$) shows a
slight decrease (increase) in the case of the hard component being
available. From the RHIC to LHC, there is a positive (negative)
correlation between $T_0$ and $\beta_T$ ($p_0$ and $n$). The
contribution of hard component slightly increases from dozens of
GeV to above 10 TeV, and it has no contribution at around 10 GeV.

In the case of considering the two components together, the mean
transverse momentum and the initial temperature increase obviously
with the increase of logarithmic collision energy in the
considered energy range. From a few GeV to above 10 TeV, the
collision system takes place possibly a transition at around 10
GeV, where the transition from a baryon-dominated to a
meson-dominated final state takes place. No matter what a
structure appears, the energy range around 10 GeV is a special one
due to the slope of $T_0$ excitation function having large
variation. Indeed, the mentioned energy range is needed further
study in the future.
\\
\\
\\
{\bf Appendix: Fit results from $AA$ collisions}
\\

Although our previous work~\cite{54} studied the excitation
functions of $T_0$ and $\beta_T$ in $AA$ collisions, different fit
functions were used there. To make a comparison with the results
from $pp$ collisions in the present work, we have to use the same
fit functions, i.e. Eqs. (7) and (8), to re-fit the spectra from
$AA$ collisions.

Figure A1 shows the spectra of $\pi^-$ and $\pi^+$ produced in
mid-rapidity range in central Cu-Cu, Au-Au, and Pb-Pb collisions
at high energies. Panels (a)--(f) represent the data by various
symbols measured by the FOPI~\cite{73}, STAR~\cite{47},
STAR~\cite{7}, PHENIX~\cite{74}, STAR~\cite{6}, and
ALICE~\cite{20} Collaborations, respectively, where for 2.24 GeV
Au-Au collisions in panel (a) only the spectrum of $\pi^-$ is
available. As the same as Fig. 1, the solid (dotted) curves are
our results for $\pi^-$ ($\pi^+$) spectra fitted by Eq. (7)
through Eqs. (1) and (3), and the dashed (dot-dashed) curves are
our results for $\pi^-$ ($\pi^+$) spectra fitted by Eq. (7)
through Eqs. (2) and (3), where only one set of parameters is
used. The values of parameters are listed in Tables A1 and A2 with
$\chi^2$ and ndof, where the values of parameters are used
directly in Fig. 2 and indirectly in Fig. 3. The fit results based
on Eq. (7) are well fitting to the experimental data measured in
central $AA$ collisions by the international
collaborations~\cite{6}\cite{7}\cite{20}\cite{47}\cite{73}\cite{74}.

Figure A2 is the same as Fig. A1, but it shows the fit results
from Eq. (8) through Eqs. (1) and (3) as well as through Eqs. (2)
and (3). As the same as Fig. 4, the solid (dotted) curves are the
results for $\pi^-$ ($\pi^+$) spectra from Eq. (8) through Eqs.
(1) and (3), and the dashed (dot-dashed) curves are the results
for $\pi^-$ ($\pi^+$) spectra from Eq. (8) through Eqs. (2) and
(3). The values of parameters are listed in Tables A3 and A4 with
$\chi^2$ and ndof, where the values of parameters are used
directly in Fig. 5 and indirectly in Fig. 6. The fit results based
on Eq. (8) are well fitting to the experimental data measured in
central $AA$ collisions by the international
collaborations~\cite{6}\cite{7}\cite{20}\cite{47}\cite{73}\cite{74}.
It should be noted that Tables A3 and A4 are nearly the same as
Tables A1 and A2 respectively, if not equal in error. The reason
is that narrow $p_T$ ranges are used, in which the first component
plays complete or main rule. The difference between Eqs. (7) and
(8) is then not obvious.
\\
\\
\\
{\bf Author Contributions:} The authors contributed to the paper
in this way: conceptualization, F.-H.L.; methodology, F.-H.L.;
software, L.-L.L.; validation, L.-L.L. and F.-H.L.; formal
analysis, L.-L.L.; investigation, L.-L.L.; resources, F.-H.L.;
data curation, L.-L.L.; writing -- original draft preparation,
L.-L.L.; writing -- review and editing, F.-H.L.; visualization,
L.-L.L.; supervision, F.-H.L.; project administration, F.-H.L.;
funding acquisition, F.-H.L.
\\
\\
\\
{\bf Acknowledgment:} Communications from Prof. Dr. Edward K.
Sarkisyan-Grinbaum are highly acknowledged.
\\
\\
\\
{\bf Funding:} This work was supported by the National Natural
Science Foundation of China under Grant Nos. 11575103 and
11947418, the Scientific and Technological Innovation Programs of
Higher Education Institutions in Shanxi (STIP) under Grant No.
201802017, the Shanxi Provincial Natural Science Foundation under
Grant No. 201901D111043, and the Fund for Shanxi ``1331 Project"
Key Subjects Construction.
\\
\\
\\
{\bf Conflict of Interest:} The authors declare that there are no
conflicts of interest regarding the publication of this paper. The
funding agencies have no role in the design of the study; in the
collection, analysis, or interpretation of the data; in the
writing of the manuscript, or in the decision to publish the
results.
\\
\\
\\
{\bf Data Availability:} The data used to support the findings of
this study are included within the article and are cited at
relevant places within the text as references.
\\
\\
\\
{\bf Compliance with Ethical Standards:} The authors declare that
they are in compliance with ethical standards regarding the
content of this paper.
\\
\\

\begin{figure*}[!htb]
\begin{center}
\includegraphics[width=16.0cm]{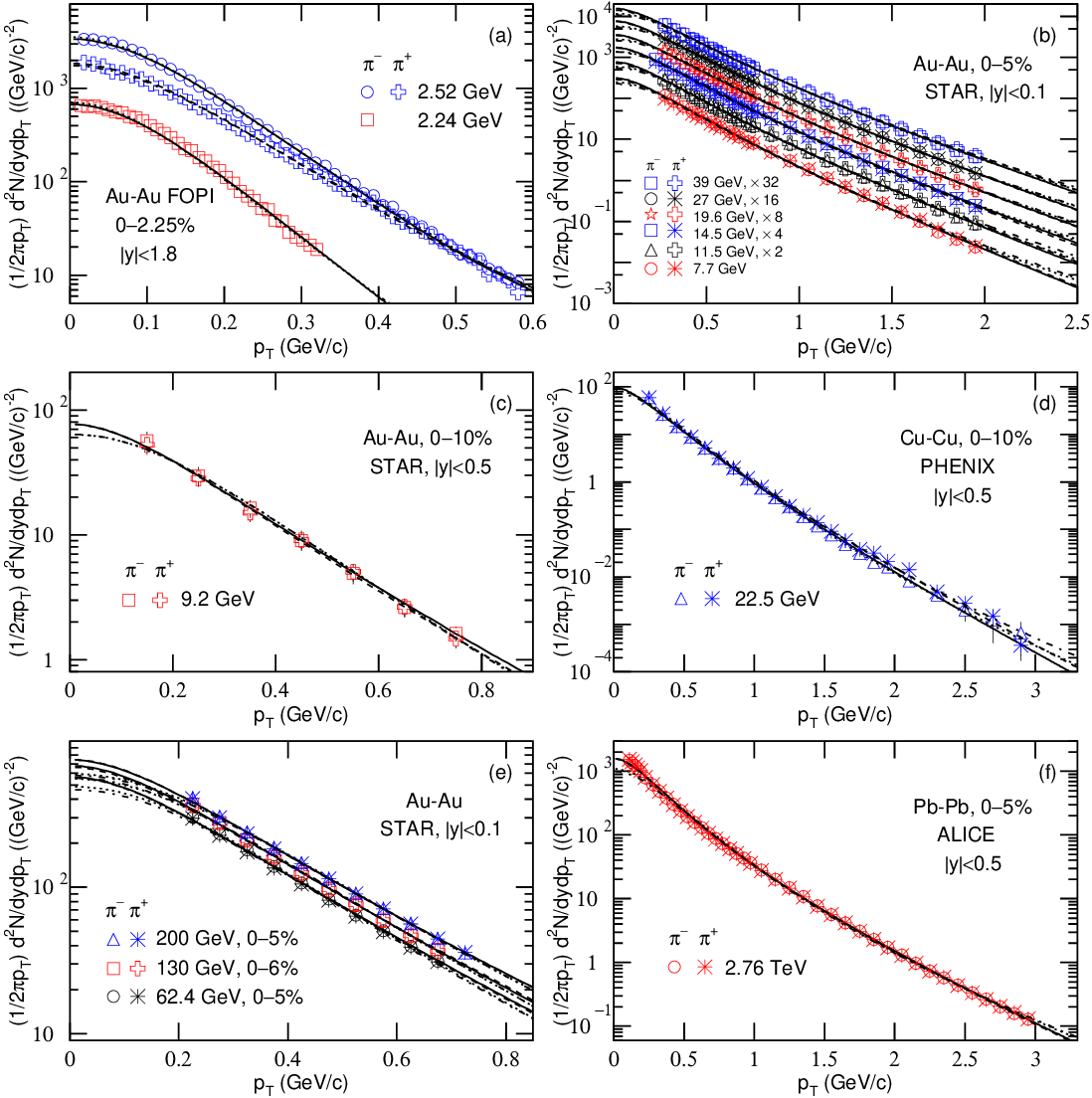}
\end{center}
{\small Fig. A1. Same as Fig. 1, but showing the results from
central $AA$ collisions. Panels (a)--(f) represent the data by
various symbols measured by the FOPI~\cite{73}, STAR~\cite{47},
STAR~\cite{7}, PHENIX~\cite{74}, STAR~\cite{6}, and
ALICE~\cite{20} Collaborations, respectively.}
\end{figure*}

\clearpage

\end{multicols}
\begin{sidewaystable}
{\scriptsize Table A1. Values of $T_0$, $\beta_T$, $k$, $p_0$,
$n$, $N_0$, $\chi^2$, and ndof corresponding to the solid (dotted)
curves for $\pi^-$ ($\pi^+$) spectra in Fig. A1.} \vspace{-4mm}
\begin{center}
{\tiny
\begin{tabular} {ccccccccccc}\\ \hline\hline Collab. & $\sqrt{s_{NN}}$ (GeV) &
Part. & $T_0$ (MeV) & $\beta_T$ ($c$) & $k$ & $p_0$ (GeV/c) & $n$ & $N_0$ & $\chi^2$ & ndof \\
\hline
FOPI Au-Au   & 2.24  & $\pi^-$ & $43\pm2$  & $0.27\pm0.02$ & 1 & $-$ & $-$ & $1734.50\pm16.33$ & 38 & 33\\
0--2.25\%    & 2.52  & $\pi^-$ & $54\pm3$  & $0.21\pm0.02$ & $0.96\pm0.01$ & $3.81\pm 0.23$ & $19.00\pm0.9$ & $2180.50\pm18.31$ & 11 & 40\\
             &  ~    & $\pi^+$ & $59\pm6$  & $0.23\pm0.02$ & $0.94\pm0.01$ & $3.91\pm0.31$  & $18.00\pm0.5$ & $1385.40\pm13.01$ & 16 & 40\\
\hline
STAR Au-Au   & 7.7   & $\pi^-$ & $118\pm6$ & $0.33\pm0.01$ & 1 & $-$ & $-$ & $20.08\pm1.08$ & 17 & 23\\
     0--5\%  &       & $\pi^+$ & $118\pm6$ & $0.34\pm0.01$ & 1 & $-$ & $-$ & $19.03\pm0.64$ & 27 & 23\\
     0--10\% & 9.2   & $\pi^-$ & $111\pm6$ & $0.34\pm0.02$ & 1 & $-$ & $-$ & $21.18\pm1.00$ & 1  & 4\\
             &       & $\pi^+$ & $110\pm6$ & $0.30\pm0.02$ & 1 & $-$ & $-$ & $21.68\pm1.62$ & 1  & 4\\
     0--5\%  & 11.5  & $\pi^-$ & $119\pm6$ & $0.35\pm0.02$ & 1 & $-$ & $-$ & $25.76\pm1.80$ & 4  & 23\\
             &       & $\pi^+$ & $120\pm6$ & $0.35\pm0.02$ & 1 & $-$ & $-$ & $24.39\pm1.32$ & 8  & 23\\
             & 14.5  & $\pi^-$ & $121\pm6$ & $0.35\pm0.02$ & 1 & $-$ & $-$ & $30.22\pm1.60$ & 1  & 25\\
             &       & $\pi^+$ & $120\pm7$ & $0.35\pm0.02$ & 1 & $-$ & $-$ & $29.47\pm1.62$ & 1  & 25\\
             & 19.6  & $\pi^-$ & $123\pm7$ & $0.36\pm0.02$ & 1 & $-$ & $-$ & $31.62\pm1.50$ & 3  & 23\\
             &       & $\pi^+$ & $124\pm7$ & $0.36\pm0.02$ & 1 & $-$ & $-$ & $31.10\pm1.52$ & 3  & 23\\
             & 27    & $\pi^-$ & $124\pm6$ & $0.36\pm0.02$ & 1 & $-$ & $-$ & $34.95\pm1.40$ & 3  & 23\\
             &       & $\pi^+$ & $124\pm5$ & $0.36\pm0.02$ & 1 & $-$ & $-$ & $34.31\pm1.22$ & 3  & 23\\
             & 39    & $\pi^-$ & $128\pm5$ & $0.36\pm0.02$ & 1 & $-$ & $-$ & $37.09\pm2.12$ & 4  & 23\\
             &       & $\pi^+$ & $129\pm5$ & $0.37\pm0.02$ & 1 & $-$ & $-$ & $35.06\pm2.42$ & 2  & 23\\
             & 62.4  & $\pi^-$ & $131\pm6$ & $0.37\pm0.02$ & 1 & $-$ & $-$ & $42.60\pm2.20$ & 9  & 4\\
             &       & $\pi^+$ & $130\pm6$ & $0.36\pm0.02$ & 1 & $-$ & $-$ & $42.16\pm2.36$ & 9  & 4\\
     0--6\%  & 130   & $\pi^-$ & $129\pm5$ & $0.37\pm0.02$ & 1 & $-$ & $-$ & $50.22\pm1.55$ & 25 & 4\\
             &       & $\pi^+$ & $130\pm6$ & $0.38\pm0.02$ & 1 & $-$ & $-$ & $48.88\pm1.09$ & 21 & 4\\
     0--5\%  & 200   & $\pi^-$ & $132\pm6$ & $0.39\pm0.02$ & 1 & $-$ & $-$ & $57.94\pm1.67$ & 6  & 5\\
             &       & $\pi^+$ & $131\pm5$ & $0.39\pm0.02$ & 1 & $-$ & $-$ & $57.81\pm1.54$ & 9  & 5\\
\hline
PHENIX Cu-Cu & 22.5  & $\pi^-$ & $127\pm6$ & $0.34\pm0.02$ & 1 & $-$ & $-$ & $35.06\pm1.85$ & 10 & 20\\
     0--10\% &       & $\pi^+$ & $126\pm6$ & $0.35\pm0.02$ & 1 & $-$ & $-$ & $34.93\pm2.21$ & 13 & 20\\
\hline
ALICE Pb-Pb  & 2760  & $\pi^-$ & $133\pm5$ & $0.42\pm0.02$ & $0.94\pm0.02$ & $5.81\pm0.34$ & $18.00\pm0.87$ & $759.82\pm12.22$ & 37 & 35\\
     0--5\%  &       & $\pi^+$ & $133\pm6$ & $0.43\pm0.02$ & $0.84\pm0.02$ & $5.79\pm0.35$ & $18.00\pm0.87$ & $754.17\pm21.00$ & 37 & 35\\
\hline
\end{tabular}}
\end{center}

{\scriptsize Table A2. Values of $T_0$, $\beta_T$, $q$, $k$,
$p_0$, $n$, $N_0$, $\chi^2$, and ndof corresponding to the dashed
(dot-dashed) curves for $\pi^-$ ($\pi^+$) spectra in Fig. A1.}
\vspace{-4mm}
\begin{center}
{\tiny
\begin{tabular} {cccccccccccc}\\ \hline\hline Collab. & $\sqrt{s_{NN}}$ (GeV) &
Part. & $T_0$ (MeV)& $\beta_T$ ($c$)& $q$ & $k$ & $p_0$ (GeV/c) & $n$ & $N_0$ & $\chi^2$ & ndof \\
\hline
FOPI Au-Au  & 2.24 & $\pi^-$ & $28\pm2$ & $0.10\pm0.02$ & $1.07\pm0.01$ & 1 & $-$ & $-$ & $1734.50\pm15.01$ & 42 & 32\\
0-2.25\%    & 2.52 & $\pi^-$ & $33\pm2$ & $0.11\pm0.02$ & $1.08\pm0.01$ & 1 & $-$ & $-$ & $2205.40\pm18.01$ & 5  & 42\\
            &  ~   & $\pi^+$ & $39\pm3$ & $0.12\pm0.02$ & $1.08\pm0.01$ & 1 & $-$ & $-$ & $1362.80\pm12.01$ & 13 & 42\\
\hline
STAR Au-Au  & 7.7  & $\pi^-$ & $88\pm6$ & $0.23\pm0.01$ & $1.06\pm0.01$ & 1 & $-$ & $-$ & $19.45\pm1.08$ & 31 & 22\\
     0--5\% &      & $\pi^+$ & $90\pm6$ & $0.24\pm0.01$ & $1.06\pm0.01$ & 1 & $-$ & $-$ & $18.64\pm0.64$ & 35 & 22\\
     0--10\%& 9.2  & $\pi^-$ & $89\pm6$ & $0.23\pm0.02$ & $1.04\pm0.01$ & 1 & $-$ & $-$ & $21.99\pm1.00$ & 2  & 3\\
            &      & $\pi^+$ & $88\pm6$ & $0.23\pm0.02$ & $1.04\pm0.01$ & 1 & $-$ & $-$ & $21.93\pm1.62$ & 2  & 3\\
     0--5\% & 11.5 & $\pi^-$ & $92\pm6$ & $0.22\pm0.02$ & $1.06\pm0.01$ & 1 & $-$ & $-$ & $25.13\pm1.80$ & 18 & 22\\
            &      & $\pi^+$ & $94\pm5$ & $0.22\pm0.02$ & $1.06\pm0.02$ & 1 & $-$ & $-$ & $24.27\pm1.22$ & 21 & 22\\
            & 14.5 & $\pi^-$ & $93\pm5$ & $0.23\pm0.02$ & $1.06\pm0.02$ & 1 & $-$ & $-$ & $30.47\pm1.20$ & 3  & 24\\
            &      & $\pi^+$ & $93\pm5$ & $0.23\pm0.02$ & $1.06\pm0.01$ & 1 & $-$ & $-$ & $29.47\pm2.22$ & 3  & 24\\
            & 19.6 & $\pi^-$ & $96\pm6$ & $0.24\pm0.02$ & $1.06\pm0.01$ & 1 & $-$ & $-$ & $32.12\pm1.30$ & 26 & 22\\
            &      & $\pi^+$ & $96\pm5$ & $0.24\pm0.02$ & $1.06\pm0.01$ & 1 & $-$ & $-$ & $30.47\pm1.22$ & 10 & 22\\
            & 27   & $\pi^-$ & $96\pm6$ & $0.24\pm0.02$ & $1.06\pm0.02$ & 1 & $-$ & $-$ & $34.32\pm2.10$ & 17 & 22\\
            &      & $\pi^+$ & $97\pm5$ & $0.25\pm0.02$ & $1.06\pm0.01$ & 1 & $-$ & $-$ & $32.79\pm1.52$ & 9  & 22\\
            & 39   & $\pi^-$ & $96\pm5$ & $0.25\pm0.02$ & $1.06\pm0.02$ & 1 & $-$ & $-$ & $35.59\pm1.42$ & 15 & 22\\
            &      & $\pi^+$ & $97\pm5$ & $0.25\pm0.02$ & $1.07\pm0.02$ & 1 & $-$ & $-$ & $34.18\pm1.22$ & 8  & 22\\
            & 62.4 & $\pi^-$ & $89\pm7$ & $0.25\pm0.02$ & $1.07\pm0.01$ & 1 & $-$ & $-$ & $41.66\pm1.20$ & 16 & 3\\
            &      & $\pi^+$ & $90\pm7$ & $0.25\pm0.02$ & $1.07\pm0.02$ & 1 & $-$ & $-$ & $40.28\pm1.62$ & 14 & 3\\
     0--6\% & 130  & $\pi^-$ & $91\pm7$ & $0.25\pm0.02$ & $1.07\pm0.01$ & 1 & $-$ & $-$ & $48.57\pm1.90$ & 57 & 3\\
            &      & $\pi^+$ & $91\pm5$ & $0.25\pm0.02$ & $1.07\pm0.01$ & 1 & $-$ & $-$ & $49.00\pm1.02$ & 49 & 3\\
     0--5\% & 200  & $\pi^-$ & $93\pm7$ & $0.26\pm0.02$ & $1.08\pm0.01$ & 1 & $-$ & $-$ & $55.93\pm1.60$ & 20 & 4\\
            &      & $\pi^+$ & $93\pm5$ & $0.26\pm0.02$ & $1.08\pm0.01$ & 1 & $-$ & $-$ & $55.10\pm1.52$ & 18 & 4\\
\hline
PHENIX Cu-Cu& 22.5 & $\pi^-$ & $94\pm6$ & $0.25\pm0.02$ & $1.05\pm0.01$ & 1 & $-$ & $-$ & $36.63\pm2.62$ & 8  & 19\\
     0--10\%&      & $\pi^+$ & $99\pm6$ & $0.25\pm0.02$ & $1.06\pm0.01$ & 1 & $-$ & $-$ & $35.94\pm3.65$ & 21 & 19\\
\hline
ALICE Pb-Pb & 2760 & $\pi^-$ & $101\pm5$& $0.33\pm0.02$ & $1.07\pm0.02$ & $0.90\pm0.02$ & $6.16\pm0.34$ & $17.93\pm0.87$ & $715.84\pm21.33$ & 88 & 34\\
     0--5\% &      & $\pi^+$ & $102\pm6$& $0.33\pm0.02$ & $1.07\pm0.02$ & $0.91\pm0.02$ & $6.26\pm0.35$ & $17.63\pm0.87$ & $671.86\pm13.56$ & 73 & 34\\
\hline
\end{tabular}}
\end{center}
\end{sidewaystable}
\begin{multicols}{2}

\clearpage

\begin{figure*}[!htb]
\begin{center}
\includegraphics[width=16.0cm]{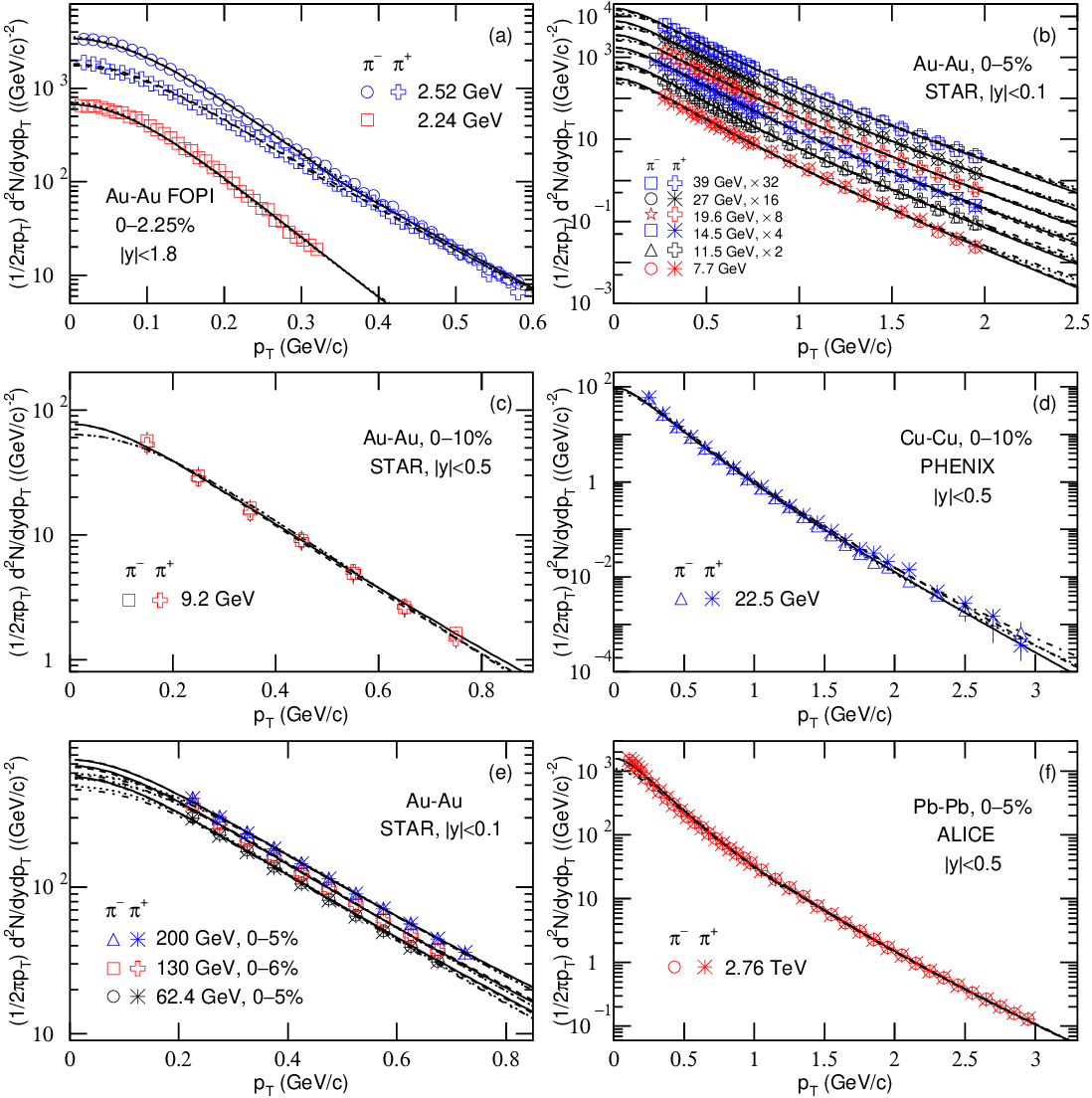}
\end{center}
{\small Fig. A2. Same as Figs. 4 and A1, but showing the results
from central $AA$ collisions and Eq. (8).}
\end{figure*}

\clearpage

\end{multicols}
\begin{sidewaystable}
{\scriptsize Table A3. Values of $T_0$, $\beta_T$, $k$, $p_0$,
$n$, $N_0$, $\chi^2$, and ndof corresponding to the solid (dotted)
curves for $\pi^-$ ($\pi^+$) spectra in Fig. A2.} \vspace{-4mm}
\begin{center}
{\tiny
\begin{tabular} {ccccccccccc}\\ \hline\hline Collab. & $\sqrt{s_{NN}}$ (GeV) &
Part. & $T_0$ (MeV) & $\beta_T$ ($c$) & $k$ & $p_0$ (GeV/c) & $n$ & $N_0$ & $\chi^2$ & ndof \\
\hline
FOPI Au-Au  & 2.24  & $\pi^-$ & $43\pm2$  & $0.27\pm0.01$ & 1 & $-$ & $-$ & $1734.50\pm15.01$ & 38 & 33\\
0--2.25\%   & 2.52  & $\pi^-$ & $54\pm3$  & $0.21\pm0.01$ & $0.96\pm0.01$ & $3.81\pm0.16$ & $19.00\pm1.19$ & $2180.50\pm22.01$ & 11 & 39\\
            &  ~    & $\pi^+$ & $59\pm3$  & $0.23\pm0.01$ & $0.94\pm0.01$ & $3.91\pm0.26$ & $18.00\pm1.25$ & $1385.40\pm33.01$ & 16 & 39\\
\hline
STAR Au-Au  & 7.7   & $\pi^-$ & $118\pm5$ & $0.33\pm0.01$ & 1 & $-$ & $-$ & $20.08\pm1.68$ & 17  & 23\\
     0--5\% &       & $\pi^+$ & $118\pm5$ & $0.34\pm0.01$ & 1 & $-$ & $-$ & $19.03\pm1.84$ & 27 & 23\\
     0--10\%& 9.2   & $\pi^-$ & $111\pm4$ & $0.34\pm0.02$ & 1 & $-$ & $-$ & $21.18\pm1.60$ & 1 & 4\\
            &       & $\pi^+$ & $110\pm5$ & $0.30\pm0.01$ & 1 & $-$ & $-$ & $21.68\pm1.62$ & 1 & 4\\
     0--5\% & 11.5  & $\pi^-$ & $119\pm6$ & $0.35\pm0.02$ & 1 & $-$ & $-$ & $25.76\pm1.45$ & 4 & 23\\
            &       & $\pi^+$ & $120\pm6$ & $0.35\pm0.02$ & 1 & $-$ & $-$ & $24.39\pm1.22$ & 8 & 23\\
            & 14.5  & $\pi^-$ & $121\pm5$ & $0.35\pm0.02$ & 1 & $-$ & $-$ & $30.22\pm1.30$ & 1 & 25\\
            &       & $\pi^+$ & $120\pm6$ & $0.35\pm0.01$ & 1 & $-$ & $-$ & $29.47\pm1.02$ & 1 & 25\\
            & 19.6  & $\pi^-$ & $123\pm6$ & $0.36\pm0.01$ & 1 & $-$ & $-$ & $31.62\pm1.15$ & 3 & 23\\
            &       & $\pi^+$ & $124\pm5$ & $0.36\pm0.01$ & 1 & $-$ & $-$ & $31.10\pm1.53$ & 3 & 23\\
            & 27    & $\pi^-$ & $124\pm6$ & $0.36\pm0.02$ & 1 & $-$ & $-$ & $34.95\pm1.80$ & 3 & 23\\
            &       & $\pi^+$ & $124\pm5$ & $0.36\pm0.01$ & 1 & $-$ & $-$ & $34.31\pm1.82$ & 3 & 23\\
            & 39    & $\pi^-$ & $128\pm7$ & $0.36\pm0.01$ & 1 & $-$ & $-$ & $37.09\pm1.22$ & 4 & 23\\
            &       & $\pi^+$ & $129\pm5$ & $0.37\pm0.01$ & 1 & $-$ & $-$ & $35.06\pm1.32$ & 2 & 23\\
            & 62.4  & $\pi^-$ & $131\pm5$ & $0.37\pm0.02$ & 1 & $-$ & $-$ & $42.60\pm1.70$ & 9 & 4\\
            &       & $\pi^+$ & $130\pm6$ & $0.36\pm0.02$ & 1 & $-$ & $-$ & $42.16\pm1.52$ & 9 & 4\\
     0--6\% & 130   & $\pi^-$ & $129\pm6$ & $0.37\pm0.01$ & 1 & $-$ & $-$ & $50.22\pm1.60$ & 25 & 4\\
            &       & $\pi^+$ & $130\pm6$ & $0.38\pm0.01$ & 1 & $-$ & $-$ & $48.88\pm1.02$ & 21 & 4\\
     0--5\% & 200   & $\pi^-$ & $132\pm5$ & $0.39\pm0.01$ & 1 & $-$ & $-$ & $57.94\pm2.20$ & 6 & 5\\
            &       & $\pi^+$ & $131\pm5$ & $0.39\pm0.02$ & 1 & $-$ & $-$ & $57.81\pm2.22$ & 9 & 5\\
\hline
PHENIX Cu-Cu& 22.5  & $\pi^-$ & $127\pm6$ & $0.34\pm0.02$ & 1 & $-$ & $-$ & $35.06\pm1.01$ & 10 & 20\\
     0--10\%&       & $\pi^+$ & $126\pm6$ & $0.35\pm0.02$ & 1 & $-$ & $-$ & $34.93\pm1.01$ & 13 & 20\\
\hline
ALICE Pb-Pb & 2760  & $\pi^-$ & $133\pm5$ & $0.42\pm0.02$ & $0.94\pm0.02$ & $5.81\pm0.34$ & $18.00\pm0.87$ & $759.82\pm14.07$ & 37 & 35\\
     0--5\% &       & $\pi^+$ & $133\pm6$ & $0.43\pm0.02$ & $0.84\pm0.02$ & $5.79\pm0.35$ & $18.00\pm0.87$ & $754.17\pm13.07$ & 37 & 35\\
\hline
\end{tabular}}
\end{center}

{\scriptsize Table A4. Values of $T_0$, $\beta_T$, $q$, $k$,
$p_0$, $n$, $N_0$, $\chi^2$, and ndof corresponding to the dashed
(dot-dashed) curves for $\pi^-$ ($\pi^+$) spectra in Fig. A2.}
\vspace{-4mm}
\begin{center}
{\tiny
\begin{tabular} {cccccccccccc}\\ \hline\hline Collab. & $\sqrt{s_{NN}}$ (GeV) &
Part. & $T_0$ (MeV) & $\beta_T$ ($c$)& $q$ & $k$ & $p_0$ (GeV/c) & $n$ & $N_0$ & $\chi^2$ & ndof \\
\hline
FOPI Au-Au  & 2.24  & $\pi^-$ & $28\pm2$ & $0.10\pm0.02$ & $1.07\pm0.01$ & 1 & $-$ & $-$ & $1734.50\pm20.31$ & 42 & 32\\
0--2.25\%   & 2.52  & $\pi^-$ & $33\pm3$ & $0.11\pm0.02$ & $1.08\pm0.01$ & 1 & $-$ & $-$ & $2205.40\pm18.12$ & 5  & 42\\
            &  ~    & $\pi^+$ & $39\pm3$ & $0.12\pm0.02$ & $1.08\pm0.01$ & 1 & $-$ & $-$ & $1362.80\pm13.22$ & 13 & 42\\
\hline
STAR Au-Au  & 7.7   & $\pi^-$ & $88\pm6$ & $0.23\pm0.01$ & $1.06\pm0.01$ & 1 & $-$ & $-$ & $19.45\pm1.58$ & 31 & 22\\
     0--5\% &       & $\pi^+$ & $90\pm5$ & $0.24\pm0.01$ & $1.06\pm0.01$ & 1 & $-$ & $-$ & $18.64\pm1.34$ & 35 & 22\\
     0--10\%& 9.2   & $\pi^-$ & $89\pm6$ & $0.23\pm0.01$ & $1.04\pm0.01$ & 1 & $-$ & $-$ & $21.99\pm1.51$ & 2  & 3\\
            &       & $\pi^+$ & $88\pm6$ & $0.23\pm0.02$ & $1.04\pm0.01$ & 1 & $-$ & $-$ & $21.93\pm2.02$ & 2  & 3\\
     0--5\% & 11.5  & $\pi^-$ & $92\pm6$ & $0.22\pm0.01$ & $1.06\pm0.01$ & 1 & $-$ & $-$ & $25.13\pm1.20$ & 18 & 22\\
            &       & $\pi^+$ & $94\pm5$ & $0.22\pm0.02$ & $1.06\pm0.01$ & 1 & $-$ & $-$ & $24.27\pm1.56$ & 21 & 22\\
            & 14.5  & $\pi^-$ & $93\pm5$ & $0.23\pm0.01$ & $1.06\pm0.01$ & 1 & $-$ & $-$ & $30.47\pm1.30$ & 3  & 24\\
            &       & $\pi^+$ & $93\pm5$ & $0.23\pm0.02$ & $1.06\pm0.01$ & 1 & $-$ & $-$ & $29.47\pm1.72$ & 3  & 24\\
            & 19.6  & $\pi^-$ & $96\pm5$ & $0.24\pm0.02$ & $1.06\pm0.02$ & 1 & $-$ & $-$ & $32.12\pm1.20$ & 26 & 22\\
            &       & $\pi^+$ & $96\pm7$ & $0.24\pm0.02$ & $1.06\pm0.02$ & 1 & $-$ & $-$ & $30.47\pm1.51$ & 10 & 22\\
            & 27    & $\pi^-$ & $96\pm7$ & $0.24\pm0.02$ & $1.06\pm0.02$ & 1 & $-$ & $-$ & $34.32\pm1.36$ & 17 & 22\\
            &       & $\pi^+$ & $97\pm7$ & $0.25\pm0.01$ & $1.06\pm0.02$ & 1 & $-$ & $-$ & $32.79\pm1.28$ & 9  & 22\\
            & 39    & $\pi^-$ & $96\pm7$ & $0.25\pm0.01$ & $1.06\pm0.01$ & 1 & $-$ & $-$ & $35.59\pm1.51$ & 15 & 22\\
            &       & $\pi^+$ & $97\pm5$ & $0.25\pm0.02$ & $1.07\pm0.01$ & 1 & $-$ & $-$ & $34.18\pm1.42$ & 8  & 22\\
            & 62.4  & $\pi^-$ & $89\pm5$ & $0.25\pm0.02$ & $1.07\pm0.02$ & 1 & $-$ & $-$ & $41.66\pm1.33$ & 16 & 3\\
            &       & $\pi^+$ & $90\pm6$ & $0.25\pm0.01$ & $1.07\pm0.01$ & 1 & $-$ & $-$ & $40.28\pm1.12$ & 14 & 3\\
     0--6\% & 130   & $\pi^-$ & $91\pm6$ & $0.25\pm0.02$ & $1.07\pm0.01$ & 1 & $-$ & $-$ & $48.57\pm1.66$ & 57 & 3\\
            &       & $\pi^+$ & $91\pm5$ & $0.25\pm0.01$ & $1.07\pm0.01$ & 1 & $-$ & $-$ & $49.00\pm1.42$ & 49 & 3\\
     0--5\% & 200   & $\pi^-$ & $93\pm6$ & $0.26\pm0.02$ & $1.08\pm0.02$ & 1 & $-$ & $-$ & $55.93\pm1.70$ & 20 & 4\\
            &       & $\pi^+$ & $93\pm5$ & $0.26\pm0.02$ & $1.08\pm0.01$ & 1 & $-$ & $-$ & $55.10\pm2.02$ & 18 & 4\\
\hline
PHENIX Cu-Cu& 22.5  & $\pi^-$ & $94\pm6$ & $0.25\pm0.01$ & $1.05\pm0.02$ & 1 & $-$ & $-$ & $36.63\pm2.66$ & 8  & 19\\
     0--10\%&       & $\pi^+$ & $99\pm6$ & $0.25\pm0.01$ & $1.06\pm0.02$ & 1 & $-$ & $-$ & $35.94\pm1.22$ & 21 & 19\\
\hline
ALICE Pb-Pb & 2760  & $\pi^-$ & $101\pm5$& $0.33\pm0.01$ & $1.07\pm0.01$ & $0.90\pm0.02$ & $6.16\pm0.34$ & $17.93\pm0.87$ & $715.84\pm11.39$ & 88 & 34\\
     0--5\% &       & $\pi^+$ & $102\pm6$& $0.33\pm0.02$ & $1.07\pm0.02$ & $0.91\pm0.02$ & $6.26\pm0.35$ & $17.63\pm0.87$ & $671.86\pm16.22$ & 73 & 34\\
\hline
\end{tabular}}
\end{center}
\end{sidewaystable}
\begin{multicols}{2}

\end{multicols}
\end{document}